\pdfoutput=0
\documentclass[useAMS,usenatbib,usegraphicx]{mn2e}
\usepackage{times,color,aas_macros}
\usepackage{url}
\newcommand{\hpix}{HEALPix}
\newcommand{\sht}{SHT}
\newcommand{\mg}{MG}
\newcommand{\shtmg}{SHT+MG}

\newcommand{\hmpc}{\ensuremath{h^{-1}\mathrm{Mpc}}}
\newcommand{\hkpc}{\ensuremath{h^{-1}\mathrm{kpc}}}

\newcommand{\sinc}{\ensuremath{{\rm sinc}}}
\newcommand{\sqrdeg}{\ensuremath{\mathrm{deg}^{2}}}
\newcommand{\calclens}{{\bf C}urved-sky gr{\bf A}vitational {\bf L}ensing for {\bf C}osmological {\bf L}ight con{\bf E} simulatio{\bf NS}}
\newcommand{\lesssim}{\la}

\title[CALCLENS: Curved-sky Weak Lensing Simulations]{CALCLENS: Weak Lensing Simulations for Large-area Sky Surveys and Second-order Effects in Cosmic Shear Power Spectra}
\author[M. R. Becker]{Matthew R. Becker\thanks{E-mail: beckermr@uchicago.edu}\\
Department of Physics, 5720 S. Ellis Avenue, The University of Chicago, Chicago, IL 60637\\
Kavli Institute for Cosmological Physics, 5640 South Ellis Avenue, The University of Chicago, Chicago, IL 60637}

\begin{document}

\date{}

\pagerange{\pageref{firstpage}--\pageref{lastpage}} \pubyear{2012}

\maketitle

\label{firstpage}

\begin{abstract}
I present a new algorithm, CALCLENS, for efficiently computing weak gravitational lensing shear signals from large N-body light cone simulations over a curved sky.  This new algorithm properly accounts for the sky curvature and boundary conditions, is able to produce redshift-dependent shear signals including corrections to the Born approximation by using multiple-plane ray tracing, and properly computes the lensed images of source galaxies in the light cone.  The key feature of this algorithm is a new, computationally efficient Poisson solver for the sphere that combines spherical harmonic transform and multgrid methods.  As a result, large areas of sky ($\sim\!10,000$ square degrees) can be ray traced efficiently at high-resolution using only a few hundred cores on widely available machines.  Using this new algorithm and curved-sky calculations that only use a slower but more accurate spherical harmonic transform Poisson solver, I study the shear B-mode and rotation mode power spectra.  Employing full-sky E/B-mode decompositions, I confirm that the shear B-mode and rotation mode power spectra are equal at high accuracy ($\lesssim1\%$), as expected from perturbation theory up to second order. Coupled with realistic galaxy populations placed in large N-body light cone simulations, this new algorithm is ideally suited for the construction of synthetic weak lensing shear catalogs to be used to test for systematic effects in data analysis procedures for upcoming large-area sky surveys.  The implementation presented in this work, written in \verb+C+ and employing widely available software libraries to maintain portability, is publicly available at \verb+http://code.google.com/p/calclens+.
\end{abstract}


\section{Introduction}\label{sec:intro}
\setcounter{footnote}{0}
Weak lensing analyses are an integral part of the scientific program of upcoming large-area sky surveys, such as the DES\footnote{The Dark Energy Survey - http://www.darkenergysurvey.org}, LSST\footnote{Large Synoptic Survey Telescope - http://www.lsst.org}, Euclid\footnote{http://sci.esa.int/euclid}, HSC\footnote{Hyper Suprime-Cam - http://www.naoj.org/Projects/HSC}, KIDS\footnote{The Kilo Degree Survey - http://kids.strw.leidenuniv.nl}, and Pan-STARRS\footnote{The Panoramic Survey Telescope \& Rapid Response System - http://pan-starrs.ifa.hawaii.edu} surveys.  They will be used to study the growth of structure via cosmic shear \citep[e.g.,][]{hoekstra2008} and will additionally serve as key followup measurements for cosmological constraints derived from the abundance of galaxy clusters as a function of mass and redshift \citep[e.g.,][]{weinberg2012}.  The importance of these observations has motivated extensive studies of all aspects of weak lensing (WL) measurements, modeling, and theory. This diverse set of topics includes the process of measuring galaxy shapes and measuring WL shear from pixelized images (e.g., STEP1/2, \citealt{heymans2006,massey2007} or GREAT08/10, \citealt{bridle2010,kitching2012}), the construction of estimators and methods to infer physically interesting quantities from catalogs of galaxy shapes and positions \citep[e.g.,][]{schneider1998,seljak1998,hu2001,jarvis2004,johnston2007a,schneider2007,schneider2010,hikage2011,becker2012}, photometric redshift estimators and their use \citep[e.g.,][]{collister2004,mandelbaum2008b,cunha2009,gerdes2010,cunha2012}, astrophysical contaminants to WL signals \citep[e.g.,][]{heavens2000,crittenden2002,hirata2004}, higher-order corrections to the first-order WL approximation itself \citep[e.g.,][]{jain2000,cooray2002,hirata2003,vale2003,dodelson2006,shapiro2006,hilbert2009,krause2010}, and the basic matter and halo statistics needed to model WL signals properly \citep[e.g.,][]{jing2006,rudd2008,hayashi2008,hilbert2010,guillet2010,lawrence2010,vandaalen2011,casarini2011,casarini2012,takahashi2012}.  

In particular, cosmological constraints from WL analyses depend sensitively on the predictions of non-linear structure formation \citep[e.g.,][]{huterer2005}. Thus both cosmological N-body (or N-body plus gas dynamic) and ray tracing simulations, which compute the shear field directly \citep[e.g.,][]{barber1999,jain2000,vale2003,hilbert2009,sato2009}, must be used to obtain even basic predictions for the statistics of WL signals \citep[e.g.,][]{semboloni2007,sato2009,sato2011,kayo2012}.  While these different types of simulations are typically used to study how non-linear structure formation impacts the analysis  of WL signals, they can also be used as the basis for synthetic WL galaxy shear catalogs for studying potential systematic effects in the process of making and interpreting WL measurements themselves \citep[e.g.,][]{vale2004,hartlap2011,heymans2012}. In the coming years, as future surveys become larger and more complicated with ever increasing statistical power, the use of simulations in this way will become increasingly important for understanding WL measurements and ensuring that the cosmological parameter constraints derived from them are free of systematic errors.  

From the point of view of WL alone, there are several requirements that these synthetic WL galaxy shear catalogs must meet.\footnote{There are of course other requirements, like realistic galaxy populations and observational effects like the confusion between stars and galaxies, but these are not the focus of this work.} Requirements directly focused on WL include computing redshift-dependent shear signals at the locations of the WL sources, properly capturing the effects of magnification on WL source sizes and magnitudes, and computing the image positions of sources.  Capturing magnification effects properly is important for testing methods to self-calibrate uncertainties in shear measurements using magnification signals \citep[e.g.,][]{rozo2010,vallinotto2011} and direct measurements of magnification signals or cross-correlations of them with other signals \citep[e.g.,][]{yang2011,heavens2011,gaztanaga2012,casaponsa2012}. Also, these WL statistics should be computed for light cone simulations which model the complete past light cone of a fiducial observer.\footnote{These light cone simulations are typically built from standard structure formation simulations either on-the-fly or as a post-processing step.  See for example \citet{evrard2002}.} Additionally, any algorithm to compute the WL signals given an observer's past light cone should properly resolve the relevant physical (i.e., a few resolution lengths of the light cone simulation) and angular scales (i.e., a few to tens of arcseconds for capturing the properties of galaxy image positions). Finally, for large-area sky surveys like those in the near future, it will be advantageous to produce shear signals for the entire sky at once, accounting properly for the sky curvature.  

This last point deserves special attention in particular.  It is certainly possible to proceed by making a large number of small-area ($\sim10^{2}$ square degree) WL simulations.  This approach has several advantages.  For a fixed dynamic range in the underlying N-body simulation, smaller area WL simulations will have better resolved structures.  Also, as I will show below, the computational methods needed for producing these simulations with a curved sky are difficult to simultaneously make fast and accurate. However, for certain scientific analyses, such as cross-correlation weak lensing measurements with optically selected galaxy clusters \citep[e.g.,][]{johnston2007,mandelbaum2008a,mandelbaum2008b}, using curved-sky pseudo-CL techniques to measure cosmic shear power spectra \citep[e.g.,][]{hikage2011}, or Cosmic Microwave Background (CMB) lensing cross-correlations with large scale structure \citep[e.g.,][]{smith2007,hirata2008,bleem2012,sherwin2012,feng2012}, large-area WL simulations on a curved-sky are preferred. At their typical densities for a survey like the DES, galaxy clusters effectively fill the sky in degree-sized patches, so that simultaneously capturing both the large- and small-angle shear signals requires high resolution over large areas with the proper treatment of the sky-curvature and boundary conditions.  Note that while angular scales below $\approx$1 arcmin may not be used for cosmic shear because of theoretical uncertainties in the matter power spectrum from galaxy formation \citep[e.g.,][]{rudd2008}, stacked weak lensing shear measurements around galaxy clusters at moderate redshifts will probe these small angular scales \citep[e.g.,][]{george2012}, motivating higher resolution calculations than those needed for cosmic shear. For optically selected galaxy clusters, the shear signals on these small scales can be biased low because of offsets between the point about which the shear is measured and the true center-of-mass or density peak of the cluster \citep[e.g.,][]{johnston2007}, exactly an effect one would like to model. Additionally, only simulations which cover the entire survey area can address potential systematics on the largest scales. Finally, as I demonstrate below, curved-sky WL simulations can be used to study higher-order WL effects directly using full-sky spherical harmonic E/B-mode decompositions, which are complete in the technical sense and free of ambiguous modes \citep[e.g.,][]{bunn2003}.\footnote{For shear or polarization fields observed only over a finite patch of sky, there exist Fourier modes in the patch which cannot be uniquely classified as E- or B-modes. These modes, called ambiguous modes in the literature, are not present when the shear or polarization field covers the entire sky.}

The method of choice for WL simulations in this context is multiple-plane ray tracing \citep[e.g.,][]{barber1999,jain2000,vale2003,forero2007,hilbert2009}.  Multiple-plane ray tracing simulations properly compute redshift dependent shear effects, second-order lensing effects \citep{dodelson2005}, all magnification effects, and can be used in conjunction with other methods, like grid search methods \citep{schneider1992}, to compute the properly lensed images of sources in the light cone \citep{hilbert2009}. To date, multiple-plane ray tracing simulations either have been implemented in the flat-sky approximation for small areas \citep[e.g.,][]{barber1999,jain2000,vale2003,hilbert2009} or account for the sky curvature properly but do not track the shear at each ray location \citep[e.g.,][]{teyssier2009}. WL simulations which employ the Born approximation in either the flat-sky or curved-sky context exist as well \citep[e.g.,][]{fosalba2008,kiessling2011}.  It is also important to note that methods used for CMB lensing simulations operate on a curved-sky and are closely related or in fact identical to those used for simulations of WL observations with galaxies, except that they typically employ either a single lens plane \citep[e.g.,][]{carbone2008,das2008,sehgal2010} or synthesize the WL deflection field assuming Gaussian statistics \citep[e.g.,][]{lewis2005}. 

In this work, I will present CALCLENS (\calclens), a curved-sky, multiple-plane ray tracing algorithm which computes the entire lensing Jacobian at the ray positions, suitable for the construction of synthetic WL galaxy shear catalogs for large-area sky surveys. In Section~\ref{sec:lensform}, I describe the multiple-plane ray tracing algorithm and how it can be extended to the sphere.  Additionally, in this section I cover the technical details of the physical approximations made when performing multiple-plane ray tracing on the sphere, showing that this algorithm, although formulated on the sphere, still works in the Limber approximation.  I also describe the method to find the images of lensed source galaxies. I described the code implementation and parallelization in Section~\ref{sec:impandpar}. In particular, I describe a new method to solve the Poisson equation on the surface of the sphere, the most time consuming step of the multiple-plane ray tracing algorithm, which combines spherical harmonic transform (SHT) and multigrid (MG) methods. In the high-resolution limit this SHT plus MG (\shtmg) method is significantly faster than pure \sht\ calculations. Note that these calculations are very sensitive to numerical errors which generate unwanted B-modes in the shear field.  While the combined \shtmg\ method presented in this work is not perfect in this regard, the numerical artifacts are well below the typical level of shape noise expected for upcoming surveys.  Thus in the context of making synthetic galaxy shear catalogs, the \shtmg\ method presented in this work is adequate and computationally efficient. Additionally, it can be used to compute the shear field over only part of the sky even more quickly. In Section~\ref{sec:codetests}, I present basic tests of the accuracy and performance of the new multiple-plane ray tracing code presented in this work. Then in Section~\ref{sec:convpowspec}, I study the power spectra of the convergence, E-mode shear, B-mode shear and rotation modes computed from the weak lensing maps constructed in Sections~\ref{sec:lensform}-\ref{sec:codetests}, verifying the relationship between the shear B-mode power spectrum and the rotation mode power spectrum expected from perturbation theory \citep{hirata2003}.  Finally, I conclude in Section~\ref{sec:conc}.

\section{Weak Lensing \& Multiple-plane Ray Tracing}\label{sec:lensform}
In this section I will describe the basic formulation of a multiple-plane ray tracing algorithm and how it can be extended from working in the flat-sky approximation \citep[e.g.,][]{barber1999,jain2000,vale2003,hilbert2009} to working with large areas on the sphere where the sky curvature matters \citep[e.g.,][]{das2008,teyssier2009}.  I will review the relevant equations and focus specifically on the approximations made to arrive at them, as opposed to deriving them in thorough detail (see for example \citeauthor{jain2000} \citeyear{jain2000} for a complete treatment). The essential physical point is that this algorithm, even when formulated on the sphere, still works in the Limber approximation (\citeauthor{limber1953}~\citeyear{limber1953}; see \citeauthor{kaiser1992}~\citeyear{kaiser1992} and \citeauthor{kaiser1998}~\citeyear{kaiser1998} for applications to weak lensing) so that the line-of-sight modes are integrated over.  Thus only modes transverse to the line-of-sight which are much shorter than the typical line-of-sight integration length will be accurately computed.  Only an algorithm which resolves modes parallel and transverse to the line-of-sight simultaneously will accurately reproduce the largest scale modes in the shear field.  Finally, after reviewing the basic formulation of the multiple-plane ray tracing algorithm, I will describe the algorithm used to find the images of lensed source galaxies in the light cone.  The details of the implementation and parallelization of these algorithms are discussed in Section~\ref{sec:impandpar}.

\subsection{Multiple-plane Ray Tracing}
My description of the weak lensing equations and the multiple-plane ray tracing algorithm follows very closely that of \citet{jain2000} and \citet{hilbert2009} \citep[see also e.g.,][]{dodelson2003,vale2003,das2008}.  The weak lensing equations can be written in the small-angle limit as 
\begin{eqnarray}\label{eqn:lighttraj}
\lefteqn{\beta(\chi) = \beta(\chi=0)}\nonumber\\
&&-\frac{2}{c^{2}}\int_{0}^{\chi}d\chi'\,\frac{r(\chi-\chi')}{r(\chi)r(\chi')}\nabla_{\beta}\Phi(\beta(\chi'),\chi')\ .
\end{eqnarray}
$\beta(\chi=0)$ is the angular position of the light ray at the observer, $\Phi$ is the gravitational potential, and $\chi$ is the comoving location of the light ray.  
The quantity $r(\chi)$ is the comoving angular diameter distance defined as 
\begin{equation}
r(\chi)=\left\{
\begin{array}{ll}
K^{-1/2}\sin(K^{1/2}\chi) & \mathrm{for\ } K>0\ ,\\
\chi & \mathrm{for\ } K=0\ ,\\
(-K)^{1/2}\sinh((-K)^{-1/2}\chi) & \mathrm{for\ } K<0\ ,
\end{array}\right.
\end{equation}
where $K$ is the spatial curvature, $K=-\Omega_{K}H_{0}^{2}/c^{2}$, $H_{0}$ is present day Hubble constant, and $\Omega_{K}$ is the present day value of the spatial curvature.  The comoving distance to the epoch with scale factor $a$ is given by 
\begin{equation}
\chi(a)=c\int_{a}^{1}\frac{da'}{{a'}^{2}H(a')}
\end{equation}
where $H(a)$ is the Hubble parameter.  For a flat $\Lambda$CDM cosmology $H(a)=H_{0}\sqrt{\Omega_{\rm m}a^{-3}+\Omega_{\rm de}}$ with $\Omega_{\rm de}=1-\Omega_{\rm m}$.  Here $\Omega_{\rm m}$ is the present epoch value of the matter density in units of the critical density and $\Omega_{\rm de}$ is a similar parameter for the dark energy. For models with time-varying dark energy, spatial curvature, or radiation, the expression for $H(a)$ is more complicated \citep[see e.g.,][]{albrecht2009}.

Note that the integral in Equation~(\ref{eqn:lighttraj}) is evaluated along the light ray's trajectory and thus is an implicit equation for $\beta(\chi)$. The gradient in this equation is defined in the small-angle limit as $\nabla_{\beta}=(\partial/\partial\beta_{1},\partial/\partial\beta_{2})$ where $(\beta_{1},\beta_{2})$ describe a coordinate system orthogonal to the light rays trajectory. Formally, the gradient in the previous equation should be evaluated transverse to the light ray's current direction of travel.  However, numerical tests have shown that using $\nabla_{\beta}$ instead causes a negligible amount of error \citep{vale2003}.  

Equations describing the evolution of the lensing Jacobian $A$ (also called the inverse magnification matrix) along the light ray's trajectory can be derived by simply taking another set of derivatives of Equation~(\ref{eqn:lighttraj}) with respect to the location of the ray at the observer $\beta(\chi=0)$ \citep[cf.][]{jain2000,hilbert2009}:
\begin{eqnarray}\label{eqn:sheartraj}
\lefteqn{A_{ij}(\chi) = \delta_{ij}}\nonumber\\
&&-\frac{2}{c^{2}}\int_{0}^{\chi}d\chi'\,\frac{r(\chi-\chi')}{r(\chi)r(\chi')}\frac{\partial^{2}\Phi(\beta(\chi'),\chi')}{\partial\beta_{i}(\chi')\partial
\beta_{k}(\chi')}A_{kj}(\chi')\ .
\end{eqnarray}
Repeated indices are summed over here and below.  The inverse magnification matrix $A$ is typically decomposed into four parts describing 
the transformations applied locally to the light rays as they propagate through the matter distribution \citep{schneider1992}
\begin{equation}\label{eqn:invmagmat}
A = \left( \begin{array}{cc}
1 - \kappa - \gamma_{1} & -\gamma_{2}+\omega \\
-\gamma_{2}-\omega  & 1 - \kappa + \gamma_{1}
\end{array}\right)\ ,
\end{equation}
where I have assumed that the rotation angle $\omega$ is small, as in \cite{vale2003}.  The parameter $\kappa$ is the convergence, $\gamma = \gamma_{1}+i\gamma_{2}$ defines the complex shear and describes the shearing of the image, and $\omega$ defines an overall rotation of the lensed image.  

In general, the shear field can be split into two types of modes, E- and B-modes, which differ in their transformation under parity (see, for example, \citeauthor{bunn2003}~\citeyear{bunn2003} for a complete description). At first order in the gravitational potential $\Phi$, where the equations above are evaluated along the unperturbed light ray trajectory, weak lensing by large scale structure only generates E-modes.  This limit is known in the literature as the Born approximation.  As studied in several previous works \citep[][]{cooray2002,hirata2003,krause2010}, deviations from the Born approximation generate B-modes in the shear field.  Additionally, it can be shown that at second order in $\Phi$ on small scales the power spectra of the rotation and the B-modes should be equal \citep{hirata2003}.  As I will show below, numerically verifying this relation is quite difficult but the simulations support this claim.  Note that at small angular scales all lensing effects at second order in the gravitational potential are captured by ray tracing simulations \citep{dodelson2005,bernardeau2010}. Finally, the coupling of the partial derivatives of $\Phi$ and the inverse magnification matrix $A$ in Equation~(\ref{eqn:sheartraj}) is known as the lens-lens coupling.  This coupling, along with the deviations from the Born approximation itself, comprise all of the second order terms in the lensing equations at small scales.

In order to derive a multiple-plane lensing algorithm, one divides the matter along the line of sight into slabs \citep[e.g.,][]{jain2000,vale2003,hilbert2009} or in the case of a spherical geometry, shells \citep[e.g.,][]{das2008,teyssier2009}.  To do this, let $\chi_{m}$ denote the comoving distance to the middle of the m-$th$ lens plane of which there are $N$ total lens planes.  Additionally, let the thickness of each lens plane be $\Delta\chi$ so that the m$th$ lens plane subtends in comoving distance from the observer the range 
\begin{displaymath}
\left(\chi_{m}-\Delta\chi/2,\chi_{m}+\Delta\chi/2\right)\equiv\left(\chi_{m-1/2},\chi_{m+1/2}\right)\ .
\end{displaymath}
Then the lensing equations can be written approximately as 
\begin{eqnarray}
\beta_{i}^{(N)}&=&\beta_{i}^{(0)}-\sum_{m=0}^{N-1}\frac{r(\chi_{N}-\chi_{m})}{r(\chi_{N})}\alpha_{i}^{(m)}\label{eqn:disclighttraj}\\
A_{ij}^{(N)}&=&\delta_{ij}-\sum_{m=0}^{N-1}\frac{r(\chi_{N}-\chi_{m})}{r(\chi_{N})}U_{ik}^{(m)}A_{kj}^{(m)}\label{eqn:discsheartraj}
\end{eqnarray}
where $\alpha_{i}^{(m)} = \partial_{\beta_{i}}\psi^{(m)}$ and $U_{ik}^{(m)}=\partial_{\beta_{i}\beta_{k}}\psi^{(m)}$.  The quantity $\psi^{(m)}$ will be called the lensing potential 
and is defined as
\begin{equation}
\psi^{(m)} \equiv \frac{1}{r(\chi_{m})}\int_{\chi_{m-1/2}}^{\chi_{m_{1/2}}}d\chi'\,\frac{2}{c^{2}}\Phi\ .
\end{equation}
This definition differs from that used by \citet{vale2003} due to how factors of the comoving angular diameter distance are factored out of Equations~(\ref{eqn:lighttraj}) and (\ref{eqn:sheartraj}), but is consistent with that in \citet{hilbert2009}.  

Numerically $\psi^{(m)}$ can be computed in the multiple-plane lensing  and Limber approximations from a two-dimensional Poisson equation 
\begin{eqnarray}\label{eqn:lenspoisson}
\nabla_{\beta}^{2}\psi^{(m)} &= & r(\chi_{m})\int_{\chi_{m-1/2}}^{\chi_{m_{1/2}}}d\chi'\,\frac{2}{c^{2}}\nabla_{\perp}^{2}\Phi\nonumber\\
& \approx & \Omega_{m}\frac{3H_{0}^{2}}{c^{2}}\frac{r(\chi_{m})}{a(\chi_{m})}\int_{x_{m-1/2}}^{\chi_{m+1/2}}d\chi'\,\delta \equiv \kappa^{(m)}\ .
\end{eqnarray}
where $\delta$ is the matter overdensity in the light cone and the symbol $\nabla^{2}_{\perp}$ is the transverse Laplacian with respect to comoving coordinates. This definition of the convergence differs by a factor of two relative to the standard definition ($\kappa_{\rm standard}=\kappa^{(m)}/2$), but is more convenient for numerical work.  As explained in detail in \citet{jain2000}, when the range of integration is the entire line of sight one can derive this last equation by integrating by parts, using the Poisson equation to relate the integral of the gravitational potential $\Phi$ along the line of sight to the matter over density, and neglecting long wavelength fluctuations along the line-of-sight via the Limber approximation.  As argued by \citet{das2008}, this approximation is formally quite poor for very thin lens planes as written above.  Also \citet{li2011} note that the above equation neglects extra terms that are non-zero when one is only considering a finite width lens plane, although for light cones constructed so that the matter is continuous along the line of sight these terms cancel when summed over the lens planes.  However, notice that the quantities of interest, namely the locations and inverse magnification matrices at the final positions of the rays in Equations~(\ref{eqn:disclighttraj}) and (\ref{eqn:discsheartraj}), are computed as sums over all of the previous lens planes.  Thus, to the extent that the corrections to the Born approximation are small, so that the partial derivatives transverse to the line of sight commute with the integral along the line of sight, the cancellation of the line-of-sight modes as required by the Limber approximation still occurs.  Finally, below I use light cones constructed continuously on the fly as the N-body simulation is running in order to avoid the effects discussed in \citet{li2011}.

Following previous authors \citep[e.g.,][]{das2008,teyssier2009}, this formalism can be extended to the sphere by promoting all of the transverse derivatives in the above equations to gradients and/or Laplacian operators on the sphere.  It is important to note that although the multiple-plane ray tracing algorithm can be formulated on the sphere in this way, the algorithm still will not resolve the largest scale fluctuations transverse to the line-of-sight properly.  The problem lies in Equation~(\ref{eqn:lenspoisson}) as described above.  A multiple-plane ray tracing algorithm neglects long-wavelength fluctuations along the line-of-sight and so cannot self-consistently resolve those same fluctuations transverse to the line-of-sight. By formulating the algorithm on the sphere in this way, one is working locally in the small angle approximation at every point on the sphere and has used the gradients and/or Laplacians on the sphere to account for boundary conditions and the rotation of the basis vectors from point to point \citep{stebbins1996}. For a survey like the DES, this approximation is expected to fail at very large angles, say $\ell\lesssim20$ \citep{loverde2008}, and at very low redshifts where the line-of-sight integration length is quite short.

\subsection{Light Ray Propagation}
Several difficulties arise when adapting Equations~(\ref{eqn:disclighttraj}) and (\ref{eqn:discsheartraj}) to the sphere. Equation~(\ref{eqn:disclighttraj}) is only correct in the small-angle approximation.  Additionally, Equation~(\ref{eqn:discsheartraj}) requires that at each step of the ray tracing algorithm one have the quantity $A_{kj}^{(m)}$ from all previous lens planes available.  This requirement is prohibitively expensive for the large-area calculations presented in this work. There are however more efficient methods which require only the $A_{kj}^{(m)}$ from the previous two lens planes \citep[e.g.,][]{vale2003,hilbert2009}.   In order to address these issues, I use a combination of the methods of  \citet{hilbert2009} and \citet{teyssier2009} to propagate the rays from lens plane to lens plane.  The results are reproduced here for completeness.  Additionally in Appendix~\ref{app:rayprop}, I demonstrate that the method I use for the lensing Jacobian, although formally derived in the flat-sky approximation by \citet{hilbert2009}, is equivalent to the relations given in Equation~(\ref{eqn:discsheartraj}) and thus appropriate for the full-sky calculations presented here.

I use the method presented in Appendix A of \citet{teyssier2009} to propagate the ray positions from plane to plane and to update their propagation directions.  In this method, the deflection angle at lens plane $m$, ${\balpha}^{(m)}\equiv\nabla \psi^{(m)}$, and the ray propagation direction before lensing ${\bbeta}^{(m-1)}$ are used to compute the new ray propagation direction ${\bbeta}^{(m)}$ through the rotation
\begin{eqnarray}\label{eqn:betaprop}
{\bbeta}^{(m)} & = & \mathbf{{\cal R}}(\mathbf{n}^{(m)}\times{\balpha}^{(m)},||{\balpha}^{(m)}||) \cdot {\bbeta}^{(m-1)}\\
{\bbeta}^{(0)} & = & \mathbf{n}^{(0)}\nonumber\ ,
\end{eqnarray}
where $\mathbf{{\cal R}}(\mathbf{n},\theta)$ is a rotation matrix which rotates vectors by an angle $\theta$ counter-clockwise about the axis $\mathbf{n}$ and $\mathbf{n}^{(m)}$ is a unit vector in radial direction at the current location of the ray on the sphere at lens plane $m$.  Using the updated propagation direction ${\bbeta}^{(m)}$, the ray's new Cartesian position $\mathbf{x}^{(m+1)}$ is given by
\begin{eqnarray}\label{eqn:posprop}
\mathbf{x}^{(m+1)} & = & \mathbf{x}^{(m)} + \lambda{\bbeta}^{(m)} \nonumber\\
\lambda^{2} + 2\lambda(\mathbf{x}^{(m)}\cdot{\bbeta}^{(m)}) + \chi_{m}^{2} - \chi_{m+1}^{2} & = & 0,\ \lambda>0\\
\mathbf{x}^{(0)} & = & \chi_{0}\mathbf{n}^{(0)}\nonumber\ .
\end{eqnarray}
This method is a purely three-dimensional, local version of Equation (\ref{eqn:disclighttraj}) and is consistent with how lensing of Cosmic Microwave Background temperature fields is done \citep[e.g.,][]{lewis2005}.

The lensing Jacobian is propagated from plane to plane via \citep[cf. Equation (16) of][] {hilbert2009}
\begin{eqnarray}\label{eqn:Atraj}
A_{ij}^{(n+1)}&=&\left(1-\frac{r(\chi_{n})}{r(\chi_{n+1})}\frac{r(\chi_{n+1}-\chi_{n-1})}{r(\chi_{n}-\chi_{n-1})}\right)A_{ij}^{(n-1)}\nonumber\\
&&+\frac{r(\chi_{n})}{r(\chi_{n+1})}\frac{r(\chi_{n+1}-\chi_{n-1})}{r(\chi_{n}-\chi_{n-1})}A_{ij}^{(n)}\nonumber\\
&&-\frac{r(\chi_{n+1}-\chi_{n})}{r(\chi_{n+1})}U_{ik}^{(n)}A_{kj}^{(n)}\\
A_{ij}^{(0)}&=&\delta_{ij}\nonumber\\
A_{ij}^{(-1)}&=&\delta_{ij}\nonumber\ .
\end{eqnarray}
I account for the change in the local tensor basis as the rays move from point to point on there sphere by parallel transporting the inverse magnification matrix (a tensor on the sphere) along the geodesic connecting connecting the old ray position with the new ray position.  This procedure is exactly that used for lensing polarization fields \citep{challinor2002} but has been adapted here for the case of general tensors on the surface of the sphere, as detailed in Appendix~\ref{app:paratrans}.

\subsection{Finding Galaxy Images}\label{sec:galimages}
Using the formalism above, a set of light rays can be propagated from the observer back to any source redshift desired.  However, what is needed is the image location and shear for the galaxies' source locations, not the source locations of the rays.  I use a grid search \citep[cf., \S3.4 of][]{hilbert2009}, as described by \citet{schneider1992}, to obtain image locations and shear values for the galaxies.  In the image plane a set of triangles is built out of the initial ray positions.  For a square grid of ray images, this construction is quite easy.  The method used to define the triangles for our grids is described below in \S\ref{sec:impandpar}.  The ray tracing defines a mapping between a triangle on the image plane and the equivalent triangle composed out of the same three rays on each source plane.  Every galaxy resides on a single source plane determined by its comoving distance from the observer.  For each galaxy on a given source plane, every triangle on that source plane is checked to see if it contains the galaxy.  If it does, the image position of the galaxy is computed using an interpolation over the values at the vertexes of the triangle.  A separate interpolation is used to compute the inverse magnification matrix at the galaxy position as described below.

While the interpolation over the triangle vertices defines an interpolation in the angular direction, the grid search as described above would assign to each galaxy the shear and image location at the middle of each lens plane.  Although the correction is small in practice, it is just as easy to propagate the rays used for finding the galaxy image positions and inverse magnification matrices from the middle of the lens plane to the galaxy's exact comoving distance, as described in Appendix~\ref{app:galradint}.  When this step is done the derivatives of the lensing potential $\psi$ are set to zero.  Note that in Equation~(\ref{eqn:posprop}) one now no longer requires $\lambda>0$ and uses the solution for the new ray position which minimizes the comoving separation between the old and new ray positions.  The radial interpolation defined here does not violate the condition that a galaxy should only be lensed by matter in front of it.  Additionally it computes the shear and location of the galaxy image using the proper lensing geometry. Using the shear at the ray locations without the radial interpolation is equivalent to computing the lensing Jacobian for the galaxy as (cf. Equation~\ref{eqn:sheartraj})
\begin{eqnarray}
\lefteqn{A_{ij}(\chi_{n}) = \delta_{ij}}\nonumber\\
&&-\frac{2}{c^{2}}\int_{0}^{\chi_{n}}d\chi'\,\frac{r(\chi_{n}-\chi')}{r(\chi_{n})r(\chi')}\frac{\partial^{2}\Phi(\beta(\chi'),\chi')}{\partial\beta_{i}(\chi')\partial
\beta_{k}(\chi')}A_{kj}(\chi')\nonumber\\
&&
\end{eqnarray}
where $\chi_{n}$ is the comoving distance to the middle of the lens plane which contains the galaxy (i.e. $\chi_{n-1/2}\leq\chi_{G}\leq\chi_{n+1/2}$ with $ \chi_{G}$ is the comoving distance to the galaxy).  Notice that the lensing kernel itself is wrong in the absence of the radial interpolation. By using the radial interpolation, I compute
\begin{eqnarray}
\lefteqn{A_{ij}(\chi_{G}) = \delta_{ij}}\nonumber\\
&&-\frac{2}{c^{2}}\int_{0}^{\chi_{G}}d\chi'\,\frac{r(\chi_{G}-\chi')}{r(\chi_{G})r(\chi')}\frac{\partial^{2}\Phi(\beta(\chi'),\chi')}{\partial\beta_{i}(\chi')\partial
\beta_{k}(\chi')}A_{kj}(\chi')\nonumber\ ,\\
&&
\end{eqnarray}
where I assume $\Phi(\beta(\chi'),\chi')=0$ for $\chi'\geq\chi_{n-1/2}$.  Thus the proper lensing geometry given a galaxy's location and the matter that lenses it (i.e. all matter within a distance $\chi_{n-1/2}$ of the observer) is enforced.  This radial interpolation is optimal at lowest order in the lens plane width in the sense that it respects the basic geometric and physical properties of weak lensing.

\section{Code Implementation and Parallelization}\label{sec:impandpar}
In the previous section I have described the equations and algorithm used for the full-sky weak lensing simulations presented in this work.  The basic steps of the algorithm are
\begin{enumerate}
\item Initialize the rays at the observer using Equations~(\ref{eqn:betaprop}) -- (\ref{eqn:Atraj}).
\item Solve Equation~(\ref{eqn:lenspoisson}), a two-dimensional Poisson equation on the sphere, for the lensing potential $\psi^{(m)}$ on the 
$m$th lens plane.
\item Compute the derivatives of $\psi^{(m)}$, $\alpha_{i}^{(m)}$ and $U_{ij}^{(m)}$, and then propagate the rays to the $m+1$th lens 
plane using Equations~(\ref{eqn:betaprop}) -- (\ref{eqn:Atraj}).
\item For galaxies within the $m+1$th lens plane, find their images and interpolate the lensing Jacobian onto them as described in Section~\ref{sec:galimages}.
\item Repeat this process until the last lens plane is traversed.
\end{enumerate}

In this section I will describe how each of these steps has been implemented in CALCLENS and how various computational issues have been addressed.  For example, the most time consuming and difficult of these steps is solving the two-dimensional Poisson equation on the sphere for the lensing potential $\psi^{(m)}$. The solution to this issue presented in this work, the \shtmg\ method, can speed these calculations up by an order of magnitude or more, at the cost of introducing extra B-modes in the shear field. Sufficiently high resolution calculations require either shared or distributed memory systems and efficient parallelization in order to be practical. The implementation presented in this work is written in \verb+C+ and parallelized with MPI\footnote{\url{http://www.mpi-forum.org}} in order to maintain portability and good performance across a variety of systems. I also make extensive use of the \hpix\footnote{\url{http://healpix.jpl.nasa.gov}} libraries \citep{gorski2005} for all steps of the computation. The rest of this section is organized as follows.  I cover the domain decomposition and layout of the rays in Section~\ref{sec:parallel}, the \shtmg\ method for solving for the lensing potential in Section~\ref{sec:shtmg}, the various interpolations needed to find the lensed galaxy images and lensing Jacobians in Section~\ref{sec:gridsearchimpl}, and the spatially indexed light cone formats for quick access to data on disk in Section~\ref{sec:finitesky}. In Section~\ref{sec:finitesky}, I also discuss how to work with light cones which do not cover the entire sphere.

\subsection{Ray Layout \& Domain Decomposition}\label{sec:parallel}
\begin{figure*}
\begin{center}
\includegraphics[scale=0.49]{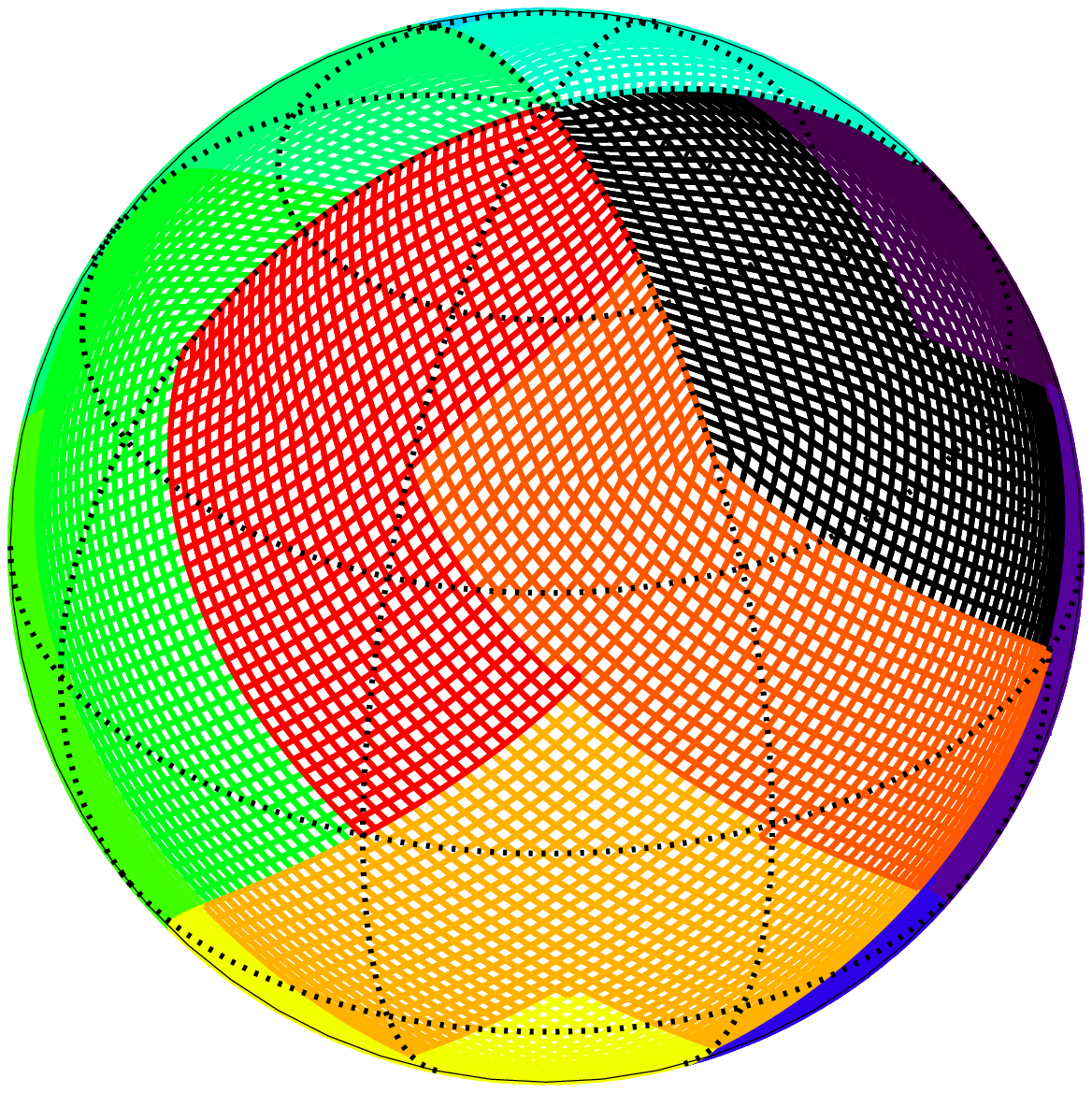}
\includegraphics[scale=0.49]{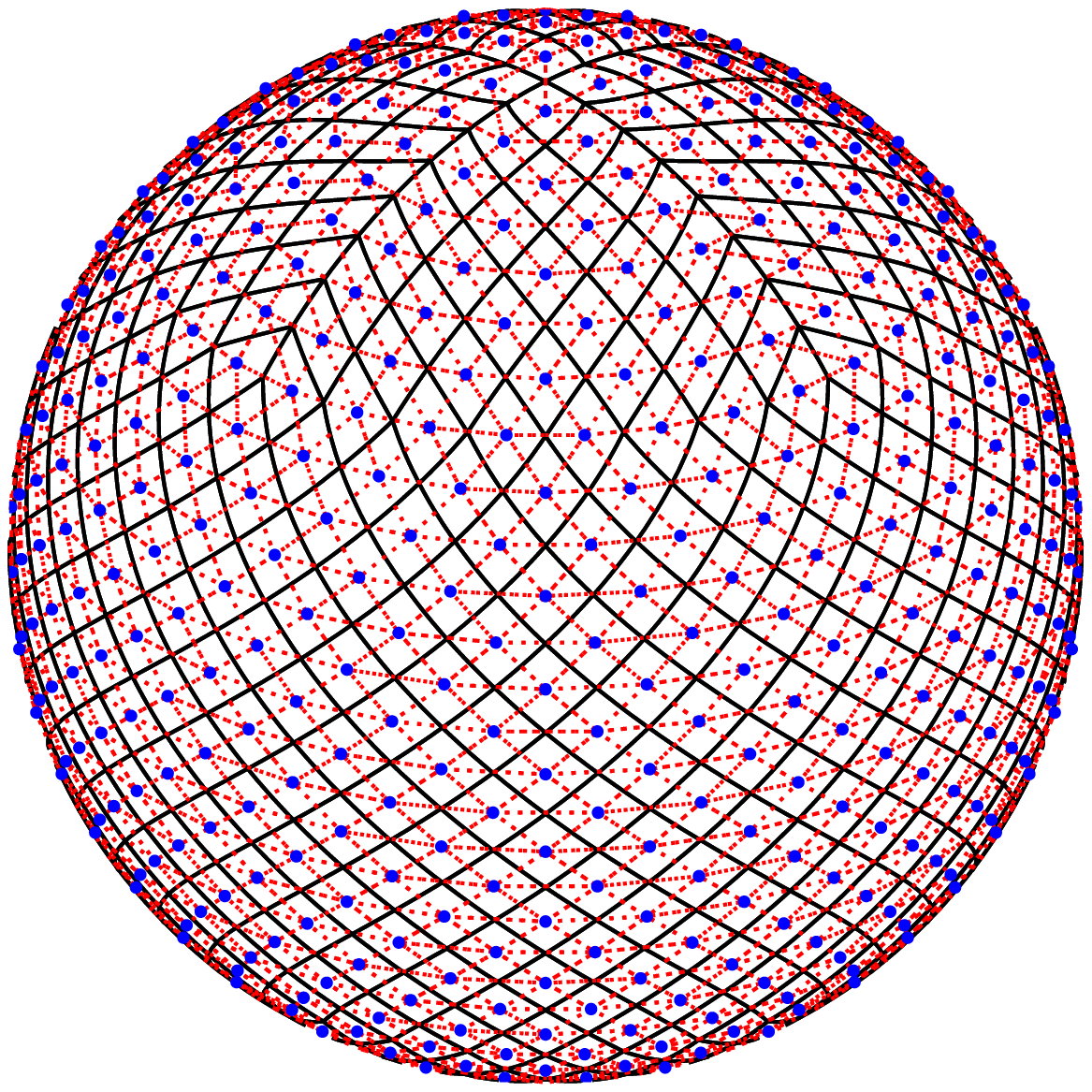}
\end{center}
\caption[]{An example domain decomposition and set of spherical triangles used for the grid search for galaxy images. The left panel shows the domain decomposition for 16 MPI tasks and ray bundles with $\mathrm{NSIDE}=32$.  Each color denotes a different domain.  The right panel shows an example set of triangles constructed out the \hpix\ cell centers for $\mathrm{NSIDE}=8$. The cell centers are shown as the blue points, the cell edges as the solid black lines, and the triangle sides are shown as the red dashed lines. Note that the triangle edges are geodesic curves whereas the \hpix\ cell boundaries are not.\label{fig:domdecgrids}}
\end{figure*}

The rays in the calculation are placed on the $z=0$ plane at the observer on \hpix\ cell centers.  In order to efficiently parallelize the computation, I then organize all of the rays into \textit{bundles} using lower resolution \hpix\ pixels, which will be denoted as bundle cells.  Using the nested indexing of the \hpix\ cells, one can compute which bundle cell each ray belongs to with quick bit shift operations.  The bundle cells with their respective rays are then distributed across processors using Peano-Hilbert indexing available in the \hpix\ package. (See Figure~\ref{fig:domdecgrids} for an example domain decomposition.)  Gravitational lensing deflections are quite weak, so that each ray moves at most a few arcmin from its original position when propagated out to redshifts of a few.  Thus I can use small buffer regions of rays and N-body particles when needed so that the rays can remain attached to the same bundle cell in which they start in order to simplify the structure of the code.  Most calculations typically employ bundles of rays with \hpix\ resolution parameter NSIDE equal to 64 or 128, producing 49,152 to 196,608 bundles covering the full sphere. Each bundle cell is approximately $0.5$ to $1$ degree across in size.  Operations like propagating the rays from lens plane to lens plane and finding the galaxy images are completely spatially local and thus embarrassingly parallel over this domain decomposition.  As I will discuss below, the most time consuming operation in the Poisson solver used in this work is also spatially local and thus embarrassingly parallel over this domain decomposition as well.

\subsection{Solving for the Lensing Potential}\label{sec:shtmg}
The Poisson equation is inherently non-local and so can be particularly difficult to solve in large parallel computations.  The  approach taken in this work is conceptually similar to that taken in the N-body community over the last few decades \citep[e.g.,][see \citeauthor{hilbert2009}~\citeyear{hilbert2009} for a specific application to weak lensing]{hockney1981,couchman1991,xu1995,kravtsov1997,bode2000,white2002b,bagla2002,springel2005}.  Namely, the solution of the Poisson equation is split into two steps.  The first step, implemented in this work with \sht s, involves tightly coupled, communication heavy computations amongst the various MPI tasks in order to solve the Poisson equation globally at low resolution. Then each MPI task uses this global solution to the Poisson equation in conjunction with some other method, a \mg\ solver in this work, to compute a higher resolution solution to the Poisson equation locally.  This last step is typically highly spatially local, so that it can be efficiently parallelized with little communication between the various MPI tasks.

The \sht\ can be used to solve the Poisson equation on the sphere by first decomposing the source term $\kappa$ into its spherical harmonic coefficients $\tilde{\kappa}_{\ell m}$ where $\ell$ and $m$ index the spherical harmonics $Y_{\ell m}(\theta,\phi)$.  This operation is typically called the {\it analysis} operation.  Then the solution to the Poisson equation is computed in harmonic space via
\begin{equation} 
-\ell(\ell+1)\tilde{\psi}_{\ell m}=\tilde{\kappa}_{\ell m}\ .
\end{equation} 
where $\tilde{\psi}_{\ell m}$ are the spherical harmonic coefficients of the lensing potential $\psi$.  The $\ell=0$ terms are set to zero by hand in this equation. Finally, the lensing potential $\psi$ is reconstituted in real space by inverting the analysis operation above.  This last step is typically called a {\it synthesis} operation. The advantage of using the \sht\ analysis and synthesis operations to solve the Poisson equation is that they automatically enforce the proper boundary conditions on the sphere.  However, the \sht\ synthesis and analysis operations have running times which scale as ${\cal O}(N^{3/2})$ with the number of resolution elements $N$ and are thus unsuitable for high-resolution calculations.  The \sht\ analysis and synthesis operations in this work are implemented over \hpix\ using fast \sht\ algorithms described in \S6.7 of \citet{press2007}.  In this work they have been parallelized with MPI in a manner similar to the \verb+S2HAT+\footnote{\url{http://www.apc.univ-paris7.fr/~radek/s2hat.html}} package.  The needed Fast Fourier Transforms are computed with the \verb+FFTW+\footnote{\url{http://www.fftw.org/}} package \citep{FFTW05}. Note that the implementation of these operations is completely parallel in memory usage as well.

Second, in order to improve on the ${\cal O}(N^{3/2})$ scaling of the \sht\ synthesis and analysis operations at high-resolution, I use the lensing potential from the \sht\ step to supply boundary conditions to a high-resolution Poisson solver which employs a finite-difference approximation to the Laplacian on the sphere and relaxation methods with \mg\ acceleration \citep{ferorenko1961,brandt1973}.  I use the full approximation scheme (FAS) to implement the \mg\ method (\citeauthor{brandt1977} \citeyear{brandt1977}, see also \citeauthor{hahn2011} \citeyear{hahn2011} for a clear explanation of the method).  The FAS scheme is implemented in this work as follows.  First for a given bundle of rays, a new coordinate system is constructed at the center of the bundle with it located at $(\theta,\phi) = (\pi/2,0)$ in the new coordinate system. Here $(\theta,\phi)$ are the angular coordinates on the sphere. The angular size of the \mg\ patches themselves is set to four times the size of the ray bundles, so that they are typically $\sim2-4$ degrees across. This coordinate system employs lines of constant $\theta$ and constant $\phi$ to define cells on the sphere with equal spacing so that $\Delta\theta=\Delta\phi$.  Near the location $(\pi/2,0)$ on the sphere, these cells have approximately the same area and aspect ratios near unity. The number of cells used at the finest level of the \mg\ patch is set so that the particle smoothing kernel, described below, is covered by $\sim4-5$ pixels and there are at least 256 pixels on a side in the patch at the finest level. 

Next, using this local pixelization and the boundary conditions from the \sht\ solution to the Poisson equation, I use red-black Gauss-Seidel relaxation with \mg\ acceleration via the FAS scheme to solve for the lensing potential. The boundary conditions are interpolated from the \hpix\ grid using the linear interpolation routine available in the public \hpix\ package. The discretization, relaxation, restriction, and prolongation methods follow that of \citet{barros1992}, except that I use direct injection instead of linear interpolation for prolongation near the boundaries to maintain stability and the averaging for the restriction operation is strictly conservative in this implementation. Note that the rays for each bundle cell are located in the center of the \mg\ patches on average. Thus by moving the boundary of the \mg\ patch, where the truncation errors are the largest, away from the rays, one can limit the effect of the truncation errors. In the center of the \mg\ patches, the \mg\ solver can in principle converge to machine precision. In practice, I use the formalism in \S20.6 of \citet{press2007} to make sure the \mg\ solver is only run until the residuals on the finest grid are slightly less than the truncation error introduced by the boundary condition interpolation scheme.  The convergence parameter, called $\epsilon$ in this work, is defined specifically as the ratio of the L2-norm of the residuals to the L2-norm of the truncation errors. Typical one uses $\epsilon\approx0.1$. The proper tuning of this convergence criterion can significantly speed up the calculations. Finally, I use the \sht\ solution of the Poisson equation to supply the starting guess for the \mg\ code.  This optimization significantly speeds up the convergence of the \mg\ routine. The accuracy and performance of the combined \shtmg\ Poisson solver is discussed below in Section~\ref{sec:codetests}.  In particular, the choice of the resolution of the \hpix\ solver relative to the \mg\ patch solver is explored in detail.

\begin{figure*}
\begin{center}
\includegraphics[scale=0.43]{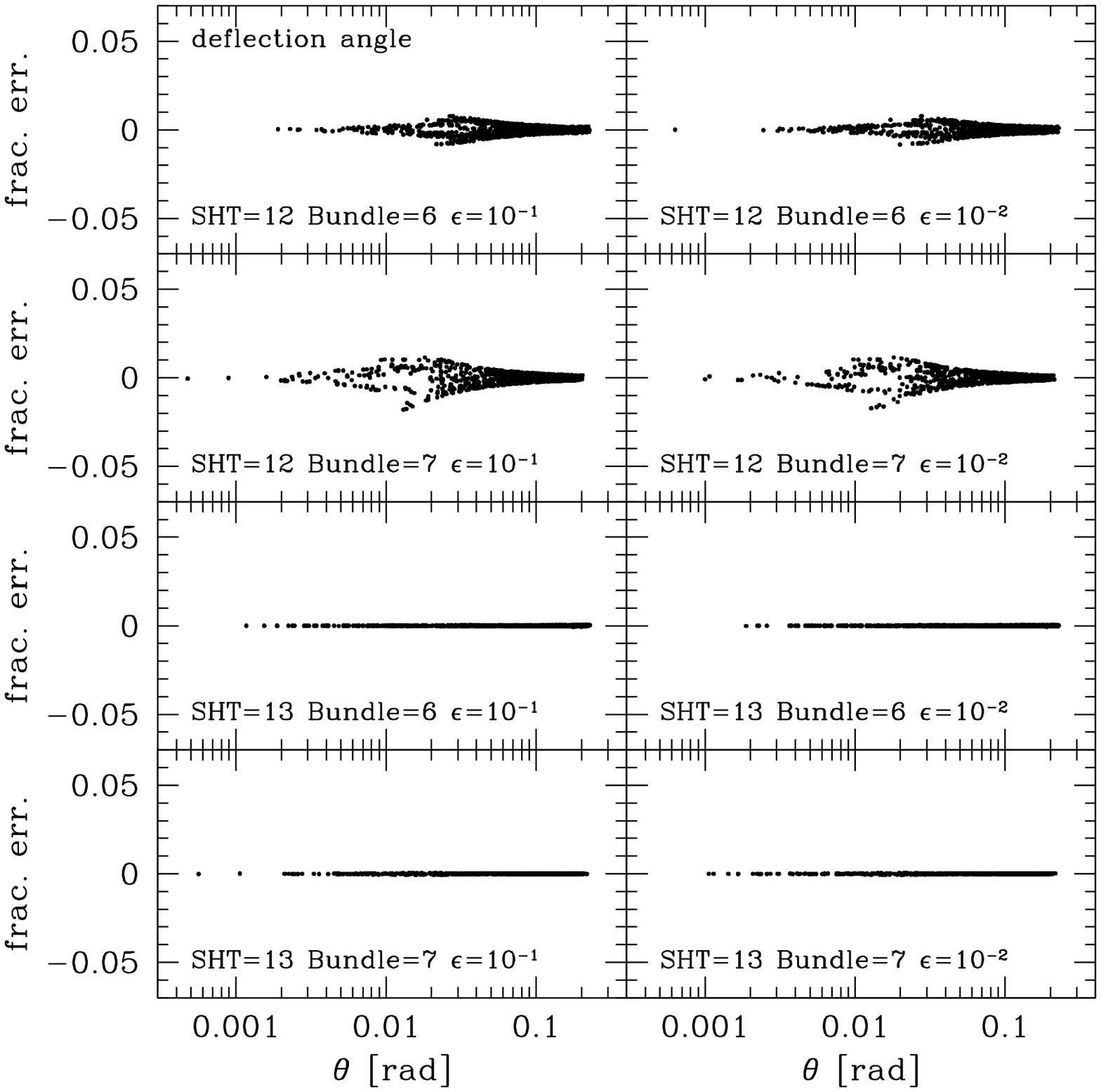}
\includegraphics[scale=0.43]{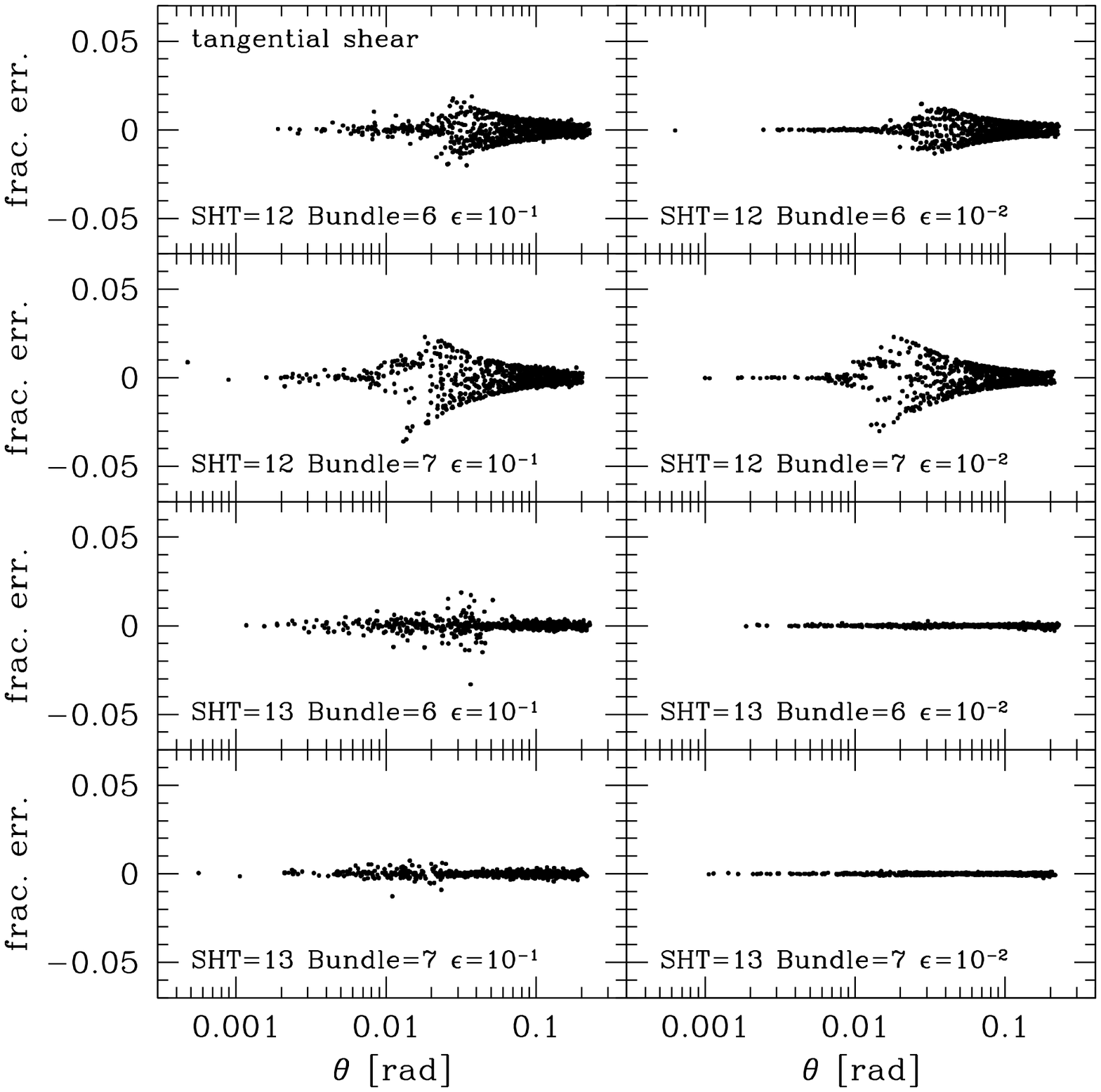}
\end{center}
\caption[]{Point mass tests of the \shtmg\ Poisson solver run in typical configurations. The left set of panels show the fractional error in the deflection angle ${\balpha}$ and the right set of panels show the fractional error in the tangential shear. Each panel corresponds to a different configuration of the \shtmg\ Poisson solver, specified by the \hpix\ order used for the \sht\ step (denoted as SHT), bundle cells (denoted as Bundle) and and the \mg\ convergence parameter $\epsilon$.  The \shtmg\ Poisson solver is unbiased at high precision ($\lesssim0.1\%$), while it has typical RMS errors of 1-2\% and has maximum absolute errors of no more than 5\% or better.\label{fig:SHTMGpmasstest}}
\end{figure*}

I use an Epanechnikov kernel \citep{epkern1969} normalized to unity on the sphere in order to adaptively smooth the mass density field sampled by the particles of the N-body simulation on each lens plane.  This kernel is \citep[cf.,][]{hilbert2009}
\begin{eqnarray}
K(\theta;\sigma) &=& \frac{1}{{\cal N}(\sigma)}\left[1 - \left(\frac{\theta}{\sigma}\right)^{2}\right]\nonumber\\
{\cal N}(\sigma) &=& 2\pi(\sinc^{2}(\sigma/2\pi) - 2\,\sinc(\sigma/\pi) +1)
\end{eqnarray}
where $\sinc(x) = \sin(\pi x)/\pi x$. The Epanechnikov kernel was chosen because it is compact and computationally efficient to implement. The smoothing length for each N-body particle is set to the larger of either a few N-body softening lengths or a few inter-ray spacings. The constraint that the smoothing length be larger than a few inter-ray spacings serves to limit the aliasing of small-scale power from the N-body particles to the \hpix\ ray grids. The exact values used are somewhat arbitrary and one typically employs $\sim2-3$ softening lengths or inter-ray spacings. The convergence field is constructed by binning the smoothing kernel over the \hpix\ map for the \sht. A similar procedure is employed for the \mg\ patches. I enforce exact mass conservation during this operation.

The partial derivatives of the lensing potential needed to propagate the rays are computed from the finest level of the \mg\ patch using fourth order finite difference stencils.  Special asymmetric stencils are used at the edge of the finest level patch (see \citeauthor{fornberg1998}~\citeyear{fornberg1998} for simple algorithms to generate these stencils) and bilinear interpolation is used to get the derivatives at the ray locations.  One must be careful to account for the non-zero Christoffel symbols for the metric on the sphere when computing these derivatives (see Appendix~\ref{app:lenspmass}).  Once the deflection angle and second-order derivatives of the lensing potential are computed at the ray locations, they are rotated back into the original coordinate system before the rays are propagated to the next lens plane.

Finally, I have also implemented a pure \sht\ algorithm which synthesizes the needed derivatives of the lensing potential directly from its spherical harmonic coefficients.  Pure \sht\ computations are slower in the high-resolution limit, but serve as good comparisons to the faster, but ultimately noisier \shtmg\ computations described above.  Also, for low-resolution shear fields, these computations are of comparable speed to the \shtmg\ computations presented here and are thus very useful. Typically one must use rays on \hpix\ grids with NSIDE = 8192 or higher for the \shtmg\ Poisson solver to be computationally more efficient than a pure \hpix\ grid. For a DES-like survey with galaxy clusters at redshift 0.8, one will need $\approx20$ arcsec effective resolution in the shear field to resolve the shear profiles at $\approx100$ \hkpc. This will require calculations with NSIDE = 16384 or higher, depending on the numerical implementation. In this regime the \shtmg\ Poisson solver will be significantly more efficient then a pure \sht\ method.

\subsection{Implementing the Grid Search}\label{sec:gridsearchimpl}
In order to implement the grid search for the galaxy images, I use \hpix\ pixel centers to construct a set of triangles which cover the sphere at the image plane.  This construction is illustrated in the right panel of Figure~\ref{fig:domdecgrids}.  I then use a fast tree code implemented in \hpix\ to search for all rays within a small radius of the source galaxy at the source plane.  This radius must be large enough to not miss any galaxy images, but small enough to be computationally efficient.  Typically, radii of a few arcmin meet these requirements.  Any triangle which has one of its vertices found near the source galaxy is tested to see if it contains the source galaxy at the source plane.  The test for whether or not the triangle contains the galaxy and the linear interpolation for the galaxy image position are implemented using barycenter coordinates as described in \S21.3 of \citeauthor{press2007} (\citeyear{press2007}; see also \citeauthor{langer2006}~\citeyear{langer2006}).  The barycenter coordinates require the triangle to be contained in a plane as opposed to the surface of the sphere.  Here I use the projection into the tangent plane to sphere computed at the galaxy source location to perform these tests.  The test for the galaxy being in the triangle is unaffected by this projection and for sufficiently small spacings between adjacent rays galaxy images are still interpolated accurately. To compute the inverse magnification matrix at the galaxy's image location, I use a separate linear interpolation over four ray image locations near the galaxy in the image plane (chosen according the interpolation routine available in the public \hpix\ package).  The lensing Jacobains are parallel transported to the galaxy image position before being used in the linear interpolation specified by the \hpix\ package.  In general, when a pure E-mode shear field is interpolated to a new a set of points, extra B-mode power can be introduced into the shear field \citep{hilbert2009}.  This second interpolation combined with the parallel transport corrections limits this effect. Finally for distributed memory computations, a buffer region of rays is imported from nearby domains so that no images are missed.  The width of this buffer region is set in principle by the maximum deflection for any photon from its observed position over its entire path.  In practice because these computations rarely resolve strong deflections and are done at moderate redshifts, a buffer region of $\sim$10 arcmin is more than sufficient.
 
\begin{table*} \begin{minipage}{100mm}
\caption{Flat $\Lambda$CDM Light Cone Simulations\label{table:sims}}
\begin{tabular}{ccccccccccc}
\hline
Simulation & Light Cone Area & Box Size & Number of Particles & Force Softening \\
\hline
Carmen & 220 \sqrdeg\ & $1000$ \hmpc\ & $1120^{3}$ & $25$ \hkpc\ \\
Lb2600 & all-sky & $2600$ \hmpc\ & $2048^{3}$ & $35$ \hkpc\ \\
\hline
\end{tabular}
\medskip
See Section~\ref{sec:codetests} for details.
\end{minipage}
\end{table*}
 
\subsection{Working with Light Cones and Finite Patches of Sky}\label{sec:finitesky}
The ray tracing algorithm is designed to work directly on N-body light cones so that it can be used easily with mock galaxy catalogs derived from the light cone data sets.  These light cone data sets are several hundred GB to many TB in size. Before the ray tracing, the N-body light cone data is organized into a spatially indexed format using \hpix\ and HDF5\footnote{{www.hdfgroup.org/HDF5/}}, so that each processor can quickly find the data it needs. In order to handle light cones which only partially cover the sphere, I set the matter density in the light cone in the regions outside the domain of interest to the mean density of the universe.  For the lensing equations presented above, rays which traverse a region of the universe with matter at its mean density will experience no lensing deflection.  This procedure limits edge effects in the shear and deflection fields for light cones with partial sky coverage. In some applications it is of interest to track rays over only part of the sphere, but still account for the influence of matter in the light cone outside of the domain of interest. For the \shtmg\ Poisson solver, this effect can be achieved by running the \sht\ step over the full sky, but running the \mg\ step only over the domain of interest.

\section{Basic Code and Scaling Tests}\label{sec:codetests}
In this section I present basic tests of the algorithm and code correctness, along with parallel scaling tests.  Section~\ref{sec:pmasstest} has tests using point masses and various configurations of the \shtmg\ Poisson solver. These tests illustrate the basic correctness of the code and the dominant source of error in the Poisson solver -- the boundary condition interpolation onto the \mg\ patches.  In Section~\ref{sec:gridsearchconvtest} I test the accuracy and convergence of the grid search implementation described above.  Finally in Section~\ref{sec:scaletests}, I present strong and weak parallel scaling tests of the code performance for the typical kinds of calculations the code is designed to perform. 

The tests presented in this section use two different N-body light cone simulations.  The first is a 220 \sqrdeg\ light cone (M. Busha \& R. Wechsler 2012, private communication) constructed from one of the Carmen simulations produced as part of the LasDamas project (C. McBride et al., in preparation)\footnote{Large Suite of Dark Matter Simulations - http://lss.phy.vander-\\bilt.edu/lasdamas}.  The Carmen simulation was run with Gadget-2 \citep{springel2001,springel2005} with second-order Lagrangian perturbation theory initial conditions \citep{crocce2006} and initial power spectrum generated with CMBFast \citep{zaldarriaga2000}.  The second is a large-volume simulation run with L-Gadget2 (an optimized version of Gadget-2 for large, dark matter only simulations), second-order Lagrangian perturbation theory initial conditions \citep{crocce2006} and initial power spectrum generated with CAMB \citep{lewis2000}. This last simulation, denoted below as Lb2600, features an all-sky light cone generated on the fly. It was generated as part of the Blind Cosmology Challenge simulation project to be described in M. Busha et al. (in preparation). Table~\ref{table:sims} lists the properties of these simulations, all flat $\Lambda$CDM models.

\subsection{Point Mass Tests}\label{sec:pmasstest}
The basic parameters which control the \shtmg\ Poisson solver are the size of the \sht\ grid, the smallest cell size used in the \mg\ step, the size of the bundle cells, and the degree of convergence required before the the \mg\ iterations are terminated.  I will denote the size of the \sht\ grid by the \hpix\ order, defined via $2^{k}\equiv\mathrm{NSIDE}$. For \hpix, each increase in order by one decreases the area of the cells by exactly a factor of four (all \hpix\ cells at a given order are equal in area) and thus the average mean inter-spacing of the cell centers by approximately a factor of two.\footnote{The area of a \hpix\ cell is exactly $\frac{4\pi}{12\times2^{2k}}$ for order $k$.}  In the following tests, the smallest \mg\ cell size is kept fixed, while the \sht\ order is varied.  Additionally, I vary the sizes of the bundle cells and the \mg\ convergence parameter $\epsilon$. This convergence parameter is proportional to ratio of the residuals to the truncation errors on the finest grid, as described above in Section~\ref{sec:shtmg}.

Fractional errors from the \shtmg\ Poisson solver using point masses and varying these parameters are shown in Figure~\ref{fig:SHTMGpmasstest}.  The left panels show the fractional errors in the deflection angle and the right panels show the fractional errors in the tangential shear.  Each panel uses a different configuration denoted by the order of the \sht\ step (12 or 13), the order of the bundle cells (6 or 7), and the convergence threshold $\epsilon$ ($10^{-1}$ or $10^{-2}$).  The smaller the parameter $\epsilon$, the higher the degree of convergence of the \mg\ algorithm. The exact choice of configuration is generally motivated by computational considerations and parallel scaling, as discussed below in Section~\ref{sec:scaletests}. The configurations shown are typical for ray tracing calculations which cover a few hundred square degrees or more.  

Several trends are apparent from this figure.  First, as the \sht\ order is increased, so that the size of the \sht\ cells approaches the size smallest \mg\ cell, the fractional errors decrease. This decrease happens because the boundary condition interpolation becomes more accurate as the \sht\ and \mg\ cell sizes become comparable.  Second, as the sizes of the bundle cells are changed, the radial location from the point mass where the fractional errors are largest shifts accordingly. These ``spiky'' features in the residuals are from bundle cells where the point mass is near the boundary of the \mg\ patches, so that the boundary condition interpolation is especially poor. Note that these errors are fixed spatially at the edges of each bundle cell, but that for a real mass distribution strong sources will be located randomly along the edges, so that these errors will tend to partially average out. These tests are not shown here, but if one increases the size of the \mg\ patches for a fixed bundle cell size and smallest \mg\ cell size, the boundary condition interpolation errors also decrease.  This decrease occurs because the \mg\ patch boundaries are moved further from the rays located at the center of the \mg\ patch. Finally one can see that when the \mg\ step is run to higher convergence by using a smaller value of $\epsilon$, the fractional errors decrease as well, until they approach the truncation errors.

From these tests it is clear that the dominant contribution to the truncation error is from the boundary condition interpolation. Overall, the \shtmg\ Poisson solver has typical RMS errors of 1-2\%, while it is unbiased at high precision ($\lesssim0.1\%$) and has maximum absolute errors of no more than 5\% or better.   As I will show below, the residual errors in the \shtmg\ Poisson solver result in extra B-mode power in the shear field, which is not present for the pure \sht\ Poisson solver.  However, these B-mode residuals are well below the level of shape noise expected for upcoming surveys so that they are tolerable in the context of making synthetic galaxy shear catalogs.

\begin{figure}
\begin{center}
\includegraphics[width=\columnwidth]{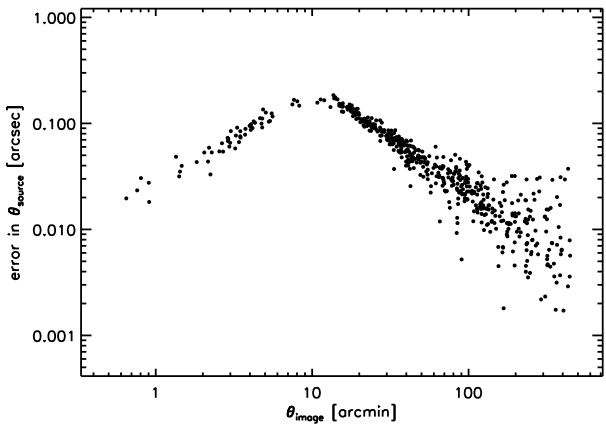}
\end{center}
\caption[]{Accuracy test of the grid search algorithm for finding weakly lensed images. Images of randomly placed background galaxies were found around a point mass smoothed with a $13.74$ arcmin smoothing length. This mass distribution does not produce multiple images. The plot shows the error in the reconstructed source locations as a function of the arc distance between the smoothed point mass and the image location. The error is defined as the arc distance between the reconstructed source location and the true source location and is a measure of the error in the lensed image locations. In the weak lensing regime, where the relationship between the image and source positions is one-to-one, the grid search has errors which are $\lesssim0.2$ arcsec. \label{fig:gspmtest}}
\end{figure}

\subsection{Tests of the Grid Search for Galaxy Images}\label{sec:gridsearchconvtest}
In this section I establish the accuracy and convergence of the grid search algorithm for sphere described above, in the weak lensing regime only.  First, I use a series of point mass tests with randomly located background sources.  For this test I use large smoothing lengths and compute the lensing properties of the smoothed mass distribution exactly (see Appendix~\ref{sec:epkern}), making sure that the smoothed mass distribution produces no strongly lensed images.  Additionally, I make sure that the distance between the source and the image positions is always less than the ray buffer region used for the grid search and also the size of the \mg\ patches.  Images due to mass distributions which exhibit lensing deflections larger than these scales will not be captured properly because the rays will either not have been imported properly from adjacent \mg\ patches or the rays themselves may not have been propagated as accurately along the line of sight because they can be deflected towards the boundaries of the \mg\ patches.  Second, I place sources randomly in a cosmological volume and test for convergence of the weak lensed image positions as the resolution of the ray grid is changed.

The results of a typical test with the smoothed point mass setup are show in Figure~\ref{fig:gspmtest}.  This figure shows the arc distance between the source and the smoothed point mass as a function of the image arc distance from the smoothed point mass. Multiple images in this figure would happen when for single source distance, multiple image horizontal lines of the same source arc distance intersect the locus of sources and images for this mass distribution multiple times. Given a computed image from the grid search, I show the reconstructed source position, computed with the lensing equations, as the blue points.  The red points show the arc distance between the reconstructed and true source positions.  In the weak lensing regime where only a single image is produced, these errors are $\lesssim0.2$ arcsec.  Although they are not shown, the errors in the shear components for the weakly lensed images are similar to those shown in Figure~\ref{fig:SHTMGpmasstest}. Finally, if the smoothing length is decreased the smoothed point mass distribution will produce strongly lensed images.  The errors in the reconstructed source positions in the strong lensing regime are typically much larger, $\sim10$ arcsec or more, increasing with the strength of the lens (but they remain roughly fixed in fractional error at a few percent).  

\begin{figure}
\includegraphics[width=\columnwidth]{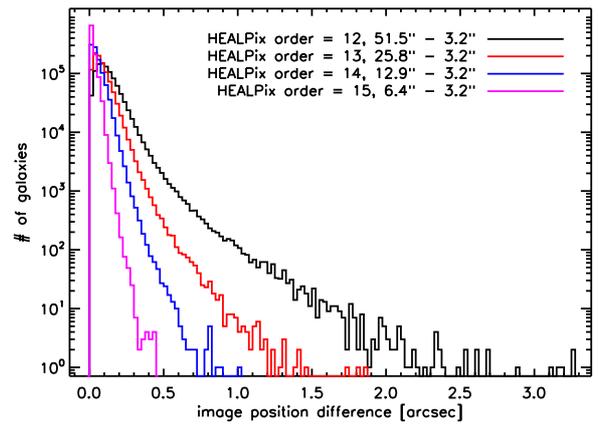}
\caption[]{Convergence test of the grid search for galaxy images. The plot shows histograms of the difference between the image positions obtained for a set of random placed sources by the grid search at different ray grid resolutions.  The lines from top to bottom show the difference between the image positions obtained with rays on a \hpix\ order 12, 13, 14, and 15 grid from those obtained with rays on a \hpix\ order 16 grid. The grid search algorithm for the sphere described above with ray grids of \hpix\ order 14 or greater have image positions converged to $\lesssim1$ arcsec for Carmen-like resolution light cones.\label{fig:gsconvtest}}
\end{figure}

Second, in order to test the convergence of the grid search algorithm implemented on the sphere, $\sim26$ million random sources were placed in the Carmen 220 deg$^{2}$ light cone.  Then the light cone was ray traced and the images of those sources computed.  The ray grid resolution for the ray tracing was varied between \hpix\ orders 12 to 16.  Figure~\ref{fig:gsconvtest} shows the difference between the image positions for the same source computed at \hpix\ order 16 and the other resolutions, \hpix\ orders 12-16. All of these images are in the weak lensing regime, so that one can expect the intrinsic accuracy of these images to be $\lesssim1$ arcsec.  Thus the convergence of the image positions in Figure~\ref{fig:gsconvtest} is largely related to changes in how the mass distribution is smoothed, which is also controlled by the ray spacing at far enough distances from the observer, as opposed to the behavior of the grid search algorithm itself.  Nevertheless, for ray grid \hpix\ orders greater than 14, the image positions are converged to $\lesssim1$ arcsec for the Carmen light cone.  This light cone has spatial and mass resolution typical of large volume simulations used for synthetic galaxy shear catalogs so that these convergence results are representative.

\subsection{Strong and Weak Scaling Tests}\label{sec:scaletests}
The ray tracing algorithm presented in this work with the \shtmg\ Poisson solver allows for versatile code implementations that have good performance in a variety of regimes.  These regimes are typically characterized by the ideas of {\it strong} and {\it weak} scaling.  In {\it strong} scaling, one measures how the running time of the computational problem changes when the size of the computational problem remains fixed (e.g., the sky area to be ray traced and resolution of the ray field) and the total number of cores participating in the calculation is varied.  Perfect parallel or strong scaling means that the running time of the problem scales inversely with the number of computational cores.  This kind of scaling is very hard to achieve in practice.  In {\it weak} scaling, one measures how the running time of the computational problem changes with the problem size per computational core fixed and the {\it total} problem size varying.  In the case of this work, one would keep the sky area per core fixed at a single resolution while varying the total sky area and thus the total number of cores.  A code with good weak scaling will have constant or decreasing total running time in this kind of test.

\begin{figure}
\includegraphics[width=\columnwidth]{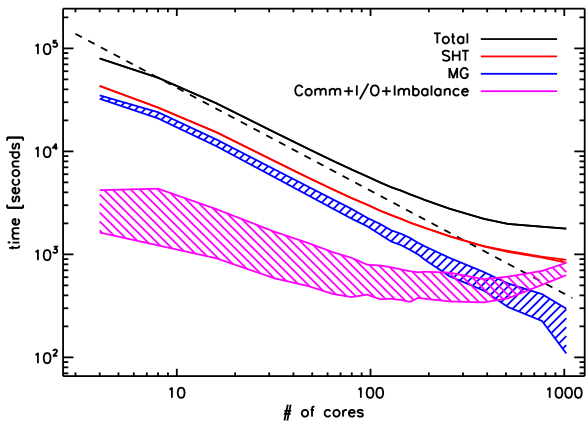}
\caption[]{Strong scaling tests of the algorithm and code for the small-area, $220$ square degree Carmen light cone. The ray tracing was done out to redshift $0.32$. The different color lines indicate the running times of different parts of the algorithm.  The hashed range shows the minimum and maximum time for a given step over all of the participating cores.  The dashed black line indicates perfect parallel scaling with the running time inversely proportional to the number of computational cores used.  Note that for ray tracing calculations with much larger areas, the running time is dominated by the \mg\ step as opposed to the \sht\ step. \label{fig:strongscale}}
\end{figure}

The implementation of the \shtmg\ algorithm presented in this work exhibits good strong scaling up to $\sim300$ cores for small area ($\sim100$ square degree) ray tracing calculations, as shown in Figure~\ref{fig:strongscale}. This figure shows the total running time and that of various computational steps for a $220$ square degree ray tracing calculation through the Carmen light cone out to redshift $0.32$.  This calculation used rays laid out on order 15 \hpix\ cells, \hpix\ order 12 \sht\ steps, and \hpix\ order 7 bundle cells.  It was performed with the Midway\footnote{\url{http://rcc.uchicago.edu/resources/midway\_specs.html}} computing cluster using Intel Xeon E5-2670 2.6 GHz processors connected via FDR-10 Infiniband.  The number of cores was varied between 4 and 1024 while the ray tracing calculation problem size remained fixed. The hatched regions in this figure show the range spanned by the minimum and maximum time for each step over all of the cores participating in the calculation.  The dashed black line shows perfect parallel scaling.  The solid black line shows the total running time, the blue hatched lines show the \mg\ step, the red lines show the \sht\ step and the magenta lines measure the rest of the running time lost to other steps in the algorithm, communication, I/O, and load balancing.  In this calculation I limited the number of processor cores which write output simultaneously to limit the load on the local file system to at most 64, which matches the scale at which the running time of this step stops scaling decently.  As the number of cores in increased, communications required to distribute \sht\ grid cells to various cores and the time required to output data to disk increase dramatically. 

For small area calculations such as this one, the overall running time is dominated by the \sht\ step.  Thus the overall parallel scaling is limited by the parallel scaling of this step.  Note that as the problem size (i.e. the area to be ray traced) is increased, the overall running time can quickly become dominated by the \mg\ step, which exhibits nearly perfect parallel scaling. Eventually, as the number of cores approaches the number of bundle cells (1629 in this case), the discreteness in the domain decomposition will limit the ability of the calculation to load balance the \mg\ step properly and thus achieve good parallel scaling.  This effect can be seen in the width of the hatched region for the \mg\ step, which increases with the number of cores indicating poorer load balancing. The discreteness in the domain decomposition, plus the overall scaling of the \sht\ step, primarily motivate the configuration of the \shtmg\ Poisson solver.  By choosing smaller bundle cells, the calculation will be able to run on more cores while still load balancing properly.  If enough cores are available, it may be advantageous to use a finer \sht\ grid in order to decrease errors in the \shtmg\ Poisson solver.  This choice comes at the expense of the running time of the \sht\ step which increases by a factor of eight for each increase in \hpix\ order.

\begin{figure}
\includegraphics[width=\columnwidth]{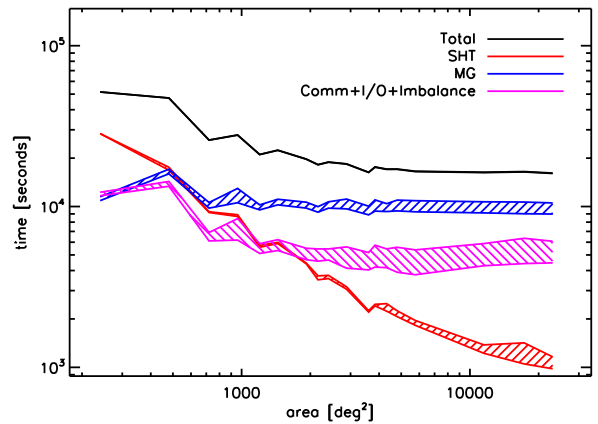}
\caption[]{Weak scaling tests of the algorithm and code. The 2600 \hmpc\ box was ray traced out to a redshift of $0.27$ using 8 cores per 214.9 \sqrdeg, varying the total number of cores between 8 and 768. This range covers 214.9 \sqrdeg\ to 20630.4 \sqrdeg\ or half of the full sky. The black, red, blue and magenta lines indicate the running times for the total calculation, the \sht\ steps, the \mg\ steps and the rest of the code.  The hatched band shows the minimum and maximum time for each step over all of the cores doing the calculation.  When the running time is dominated by the \mg\ steps, the calculations exhibit good weak scaling. \label{fig:weakscale}}
\end{figure}

In order to test the weak scaling of the code and algorithm, I compute the shear field for the Lb2600 light cone out to a redshift of 0.27, using 8 cores per 214.9 \sqrdeg.  The code was run in the same setup as the strong scaling tests using the same machine.  The results for varying the number of cores between 8 and 768, and thus the area between 214.9 \sqrdeg\ to 20630.4 \sqrdeg\ (or half of the full-sky), are shown in Figure~\ref{fig:weakscale}.  The total running time, that of the \sht\ step, that of the \mg\ step and the rest of the running time are shown in the black, red, blue and magenta lines respectively.  Again the hatched regions show the minimum and maximum time over all processors.  Once the running time of the calculation is dominated by the \mg\ step, the code exhibits good weak scaling.  These results indicate that a full calculation out $z\approx2$ or so (approximately 5 times as many lens planes) with the same setup for $\approx10,000$ \sqrdeg\ on the same machine should take $\approx10$k CPU hours, or approximately a day on 384 cores. 

\begin{figure*}
\begin{center}
\includegraphics[scale=0.43]{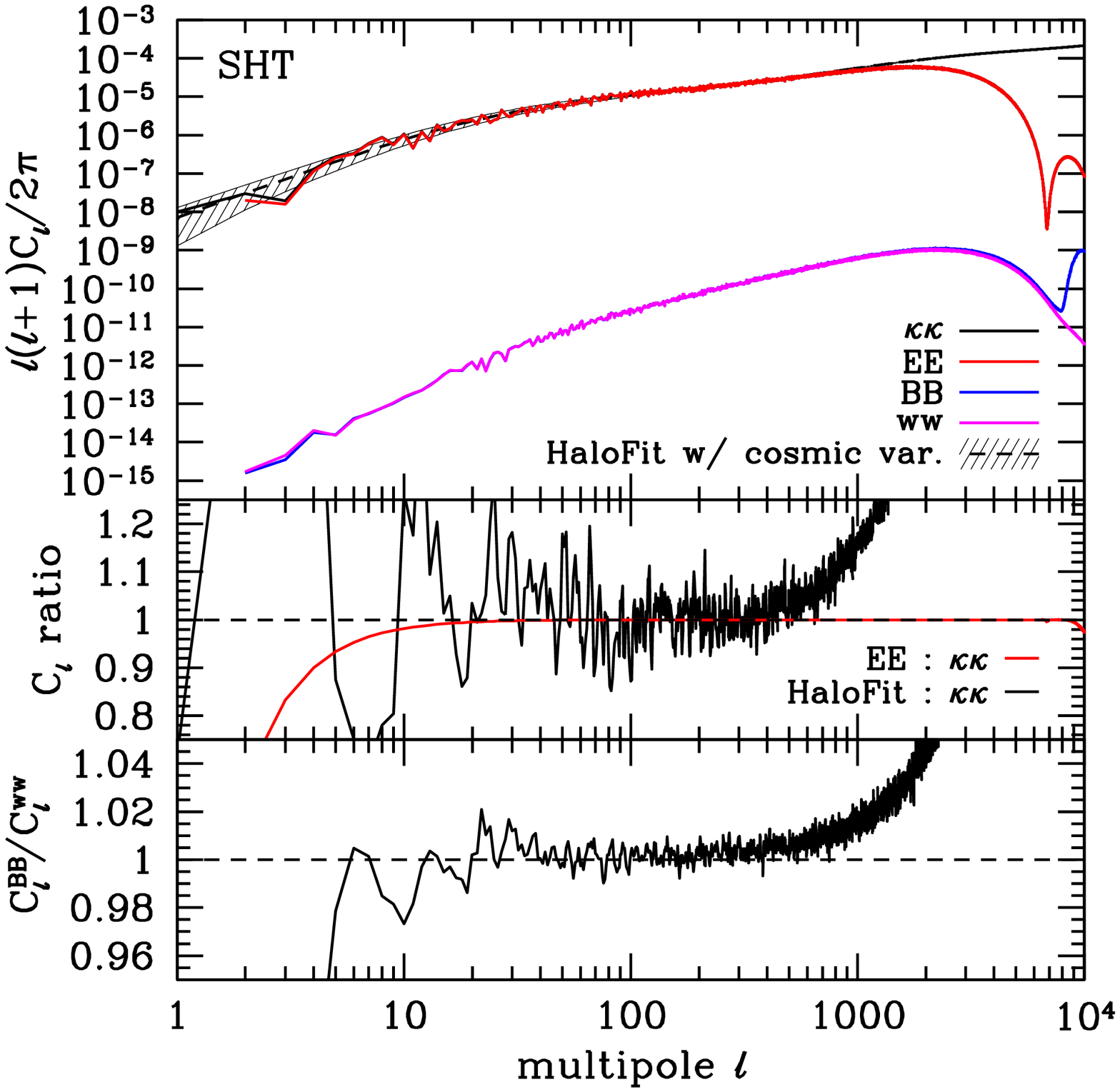}
\includegraphics[scale=0.43]{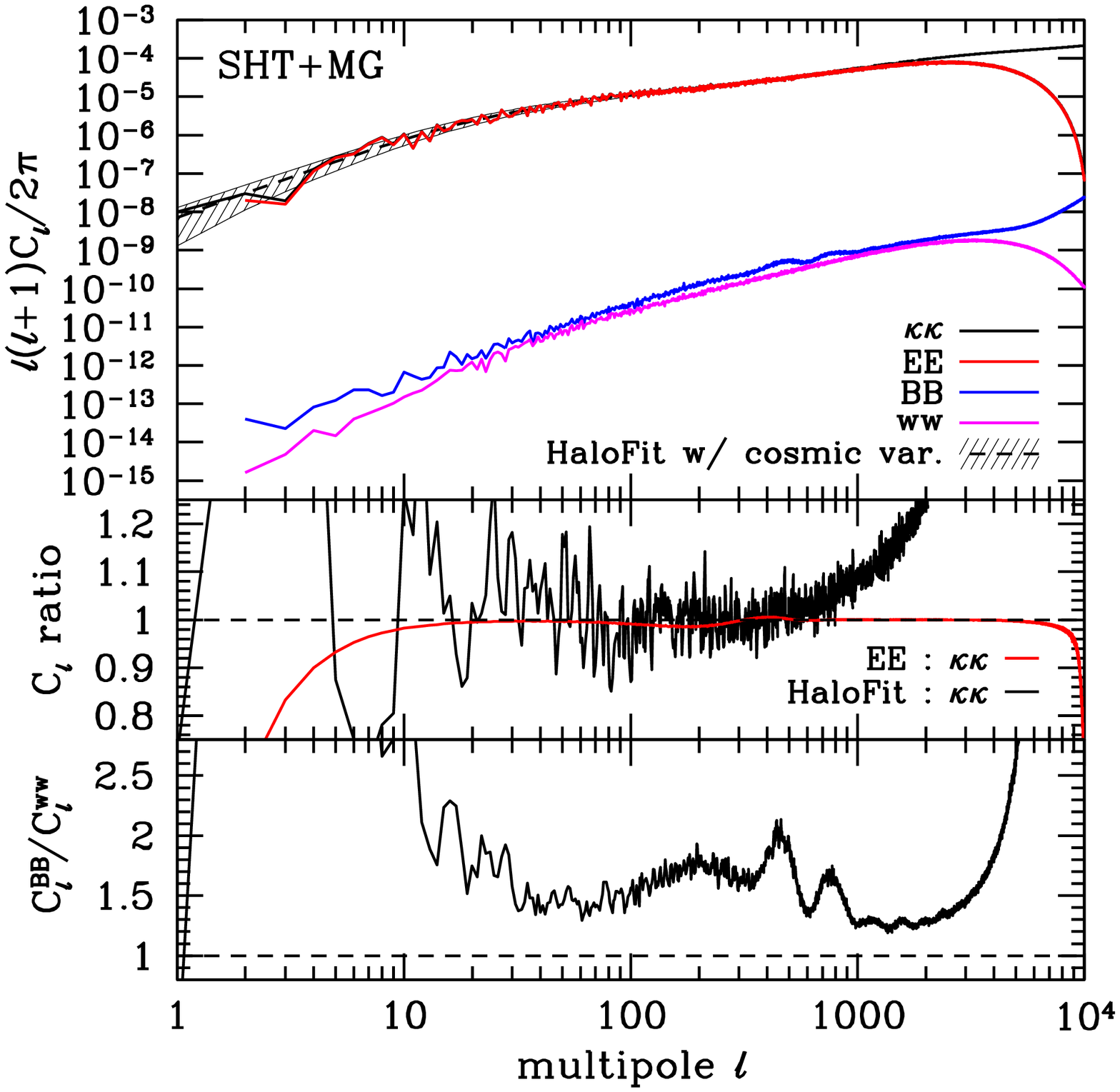}
\end{center}
\caption[]{The convergence, shear, and rotation power spectra for the pure \sht\ (left) and \shtmg\ (right) Poisson solvers at $z=1.1$. In the top panels, the black solid lines are the convergence power spectrum, the red solids are the shear E-mode power spectrum, the blue solid lines are the shear B-mode power spectrum and the magenta solid lines are the power spectrum of the rotation angle, $w$.  The dashed black line with hatched band shows the non-linear power spectrum prediction for the shear E-mode power using the revised HaloFit fitting formula of \citet{takahashi2012} with N-body shoot noise \citep{vale2003} and cosmic variance calculated assuming Gaussian statistics.  The middle panels show with the red line the ratio of the E-mode shear power to the convergence power and with the black line the ratio of the non-linear power spectrum prediction (HaloFit) to convergence power.  The bottom panels show the ratio of the shear B-mode power to the rotation power.  Both the ratio of the non-linear power spectrum prediction to the convergence power spectrum and the ratio of the B-mode power to the rotation power have been smoothed in these panels. The deviations at $\ell\ga500$ are do to the resolution of the ray field used to measure the power spectra and smoothing applied to the mass distribution, not the underlying N-body simulation.\label{fig:powspec}}
\end{figure*}

\section{The Convergence, Shear, and Rotation Power Spectra}\label{sec:convpowspec}
In this section I present the convergence, shear, and rotation power spectra for full-sky multiple-plane ray tracing calculations through N-body light cone simulations.  The Lb2600 light cone was ray traced out to a comoving distance of 2600 \hmpc\ using a \hpix\ grid of rays with \hpix\ order 12. Note that this light cone has replications in it, but they are treated self-consistently by the light cone generator (M. Busha et al., in preparation) and the ray tracing code.\footnote{The light cone generator of M. Busha et al. (in preparation) places a single observer in the periodic lattice formed by the computational volume and its replications. From this single observer, a light cone is built as the simulation is running by looking for particles which cross the light cone surface in all directions in a sufficient set of replications around the observer to capture all matter out to some desired redshift. Thus all of the boundaries are continuous and the replications are treated self-consistently by both the N-body code and the ray tracing code.} The pure \sht\ Poisson solver was run with \hpix\ order 12. The \shtmg\ Poisson solver was run with \hpix\ order 7 bundle cells and an \hpix\ order 9 \sht\ grid.  The resulting convergence, shear and rotation fields were then extracted from the inverse magnification matrices at the final plane using Equation~(\ref{eqn:invmagmat}).  Finally, the convergence, E-, B- and rotation mode power spectra were extracted from the \hpix\ maps using the public \hpix\ package.  Note that because the maps cover the entire sky, there is no ambiguity in the E/B-mode decomposition of the map \citep{bunn2003}.

The results from this simulation are shown in Figure~\ref{fig:powspec}.  The left panel shows the power spectra from the simulation using a pure \sht\ Poisson solver whereas the right panel shows the same power spectra using the \shtmg\ Poisson solver.  The top panels in this figure show the convergence, shear E-mode, shear B-mode, and rotation angle power spectra.  Additionally, the prediction in the Limber approximation for the convergence power spectrum is shown by the dashed line, computed with the revised HaloFit non-linear fitting formula of \citet{takahashi2012} and the linear power spectrum fitting formula of \citet{eisenstein1998}.  I have added the contribution of N-body shot noise to this prediction as described in \citet{vale2003}. The hatched region shows the range of expected cosmic variance assuming Gaussian statistics, a decent approximation in the linear regime \citep[see, e.g.,][]{sato2009}. The middle panels show the ratio of the non-linear prediction and the E-mode shear power spectrum to the measured convergence power spectrum. The large deviations of the power spectra from the theoretical predictions at $\ell\ga500$ are due to the smoothing applied to the mass distribution and resolution of the ray field, not the underlying N-body simulation. I have confirmed that the \citet{takahashi2012} fitting formula reproduces the simulation results more accurately than the \citet{smith2003} formula, which is low by a few percent at these scales as noted previously \citep[e.g.,][]{hilbert2009,eifler2011,takahashi2012}.

As expected at second order in General Relativity, in both cases the shear E-mode and the convergence power spectra are equal to high precision. For the \shtmg\ Poisson solver there is a slight modulation of the shear E-mode power on the scale of a few degrees, presumably arising from the boundary condition interpolation for the \mg\ patches.  However this numerical artifact in amplitude is $\lesssim2\%$, so that it will be undetectable for a DES-like survey (a signal-to-noise of $\lesssim 1$) and only marginally so for a LSST-like survey (a signal-to-noise of $\lesssim 2$). These signal-to-noise calculations were done using typical parameters for the lensing source density (12 and 40 galaxies per arcmin$^{2}$ for DES and LSST respectively) and survey area ($1/8$ and $1/2$ of the sky for the DES and LSST respectively) assuming Gaussian statistics at a variety of redshifts \citep[see e.g.,][]{huterer2005}. In both cases the B-mode and rotation mode power spectra are suppressed by nearly a factor of $10^{6}$, though this number is redshift dependent.  These results are qualitatively consistent with those already in the literature for the amplitude of the B-mode and rotation mode power spectra \citep[e.g.,][]{jain2000,vale2003,hilbert2009}.  

For the first time, one can directly compare the amplitude of the B-mode and rotation mode power spectra from the ray tracing simulations.  These are expected to be equal at second order in the gravitational potential from perturbation theory \citep[][see also \citeauthor{krause2010}~\citeyear{krause2010}]{hirata2003}.  The bottom panels of Figure~\ref{fig:powspec} show the ratio of the shear B-mode power to the rotation power.  While the pure \sht\ simulations are consistent with this result to a percent or less, the \shtmg\ simulations have a significant amount of extra B-mode power in the shear field.  This extra numerical B-mode power is roughly fixed in amplitude while the true B-mode and rotation power increase with redshift. Thus at lower redshifts, the B-mode power can be a factor of $\sim10$ above the rotation mode power.  As discussed above, this extra power arrises from the boundary condition interpolation.  Note however that this extra power is well below the expected level of shape noise for upcoming weak lensing surveys so that it is undetectable when shape noise is included (a signal-to-noise $\ll1$ for both the DES and LSST). Thus, while the \shtmg\ Poisson solver does not allow the ray tracing calculations to work completely correctly at second order, for future surveys in the context of making synthetic catalogs, these errors are acceptable.  Finally, note that given the sensitivity of the amplitude of the B-mode power to the numerical details of the simulations \citep[see also][]{hilbert2009}, I have made no attempt to compare the B-mode power spectrum amplitude in the simulations to that expected from perturbation theory as done in \citet{hilbert2009}.

\section{Conclusions}\label{sec:conc}
In this work I have presented CALCLENS, a new curved-sky ray tracing algorithm and code appropriate for current and future large-area sky surveys.  The key feature of this code is the new \shtmg\ Poisson solver for the sphere.  While this Poisson solver has more noise than pure \sht\ Poisson solvers, it enables large areas to be ray traced quickly on a curved sky at high resolution.  Additionally, this extra noise is well below the expected level of shape noise for current and upcoming surveys.  Thus this new code is ideally suited for the generation of full-sky synthetic galaxy shear catalogs for verifying the accuracy of cosmological constraints from upcoming surveys.  These synthetic galaxy shear catalogs can also be used as inputs to image simulations in order to test the process of producing cosmological constraints with weak lensing completely from images to error bars.  

I have also investigated second-order lensing effects using full-sky E/B-mode decompositions which are complete and free of ambiguous modes in the shear field \citep{bunn2003}.  Using slower, but more accurate pure \sht\ calculations I have verified that the B-mode shear and rotation mode power are equal to high precision ($\lesssim1\%$), as predicted from perturbation theory \citep{hirata2003}.  I find that the updated HaloFit prescription for the non-linear power spectrum presented by \citet{takahashi2012} used in the Limber approximation match the amplitude of the convergence and shear power spectra from ray tracing to a few percent, whereas the \citet{smith2003} prescription does not.

The calculations and methods presented in this work can be extended in several ways.  Upcoming, high-sensitivity and high-angular resolution observations of CMB temperature and polarization signals will allow for high-significance CMB lensing measurements \citep[e.g.,][]{seljak2000b,smith2006b,deputter2009,mcmahon2009,niemack2010}.  These CMB lensing maps can be cross-correlated with observations of galaxies from surveys to extract constraints on galaxy bias and other cosmological parameters \citep[e.g.,][]{deputter2009,sherwin2012,bleem2012}. Thus producing lensed CMB temperature and polarization maps for the same underlying large scale structure in which the galaxies have been placed self-consistently will be important for testing and understanding these methods.  In fact such simulations for the 220 \sqrdeg\ Carmen light cone have already been made and used in \citet{bleem2012} with data from the SPT\footnote{South Pole Telescope - \url{http://pole.uchicago.edu}} to study the galaxy bias of galaxies detected in the Blanco Cosmology Survey, similar to studies to be performed with the combination of data from the SPT and the DES.  Lensed CMB signals may be useful to study cluster mass profiles as well \citep{seljak2000}.

Current and future large-area sky surveys will present an unprecedented data analysis challenge for WL studies, both in terms of the measurement of shear signals from pixelized images and also in terms of the modeling required to reliably extract cosmological information from these measurements. In this work I have presented a new algorithm and code capable of producing full-sky weak lensing shear simulations, suitable for testing data analysis procedures and verifying cosmological constraints from these surveys are free of systematic errors. Combined with methods that place galaxies self-consistently into cosmological light cone simulations (e.g., GALACTICUS, \citealt{benson2012} or ADDGALS, R. Wechsler et al., in preparation; M. Busha et al., in preparation), this code can be used to make synthetic galaxy shear catalogs for use with a variety of current and future surveys.  As the amount and quality of WL data increases, I expect these shear catalogs, and the algorithms needed to produce them, to be increasingly useful and necessary in order to realize the scientific potential of the next generation of large-area sky surveys.  

\section*{Acknowledgments}

I am grateful to Andrey Kravtsov, Tom Abel, Eduardo Rozo, Wayne Hu, Risa Wechsler, Michael Busha, Gus Evrard, Doug Rudd and the DES weak lensing and cluster working groups for their comments and suggestions during this work. I am also grateful to those who helped debug and test the algorithms and code presented in this work by using early versions of the shear catalogs, especially J\"{o}rg Dietrich, Tim Eifler, Will High, and Alexie Leauthaud. I am additionally grateful to Volker Springel for publicly releasing Gagdet-2, upon which some of the parallel communication schemes used in this work were based. I am finally grateful to Stuart Marshall and Doug Rudd for their assistance with the Orange and Midway compute clusters respectively.  I would like to specifically thank Michael Busha, Risa Wechsler, and the LasDamas team for the use of the Carmen simulation and light cone. I also thank Brandon Erickson for running the the Lb2600 simulation used in this work.  Some of the results in this paper have been derived using the \hpix\ \citep{gorski2005} package. 

This work was supported in part by the Kavli Institute for Cosmological Physics at the University of Chicago through grants NSF PHY-0551142, NSF PHY-1125897, and an endowment from the Kavli Foundation and its founder Fred Kavli.  Some of the code development done in work was performed on the Joint Fermilab - KICP Supercomputing Cluster, supported by grants from Fermilab, Kavli Institute for Cosmological Physics, and the University of Chicago. This work used the Extreme Science and Engineering Discovery Environment (XSEDE), which is supported by National Science Foundation grant number OCI-1053575. The simulations used in this work were run on the Orange cluster at SLAC, the Ranger cluster at TACC, and the Midway cluster at the University of Chicago Research Computing Center. This work made extensive use of the NASA Astrophysics Data System and {\tt arXiv.org} preprint server.

\appendix

\section{Weak Gravitational Lensing by Point Masses on the Sphere}\label{app:lenspmass}
In this Appendix I review differential geometry on the sphere and present the solution for the shear and deflection angles due to a point mass on the sphere. On the sphere the metric is
\begin{displaymath}
ds^{2}=d\theta^{2}+\sin^{2}\theta\,d\phi^{2}
\end{displaymath}
and the non-zero Christoffel symbols are
\begin{eqnarray}
\Gamma^{\theta}_{\phi\phi}&=&-\sin\theta\,\cos\theta\nonumber\\
\Gamma^{\phi}_{\theta\phi}&=&\frac{\cos\theta}{\sin\theta}\nonumber\ .
\end{eqnarray}
Given the metric I can also define the following basis vectors
\begin{eqnarray}
\hat{\mathbf{e}}_{\theta}&=&\hat{{\bf \theta}}\nonumber\\
\hat{\mathbf{e}}_{\phi}&=&\sin\theta\,\hat{{\bf \phi}}\nonumber\\
\hat{\mathbf{e}}^{\theta}&=&\hat{{\bf \theta}}\nonumber\\
\hat{\mathbf{e}}^{\phi}&=&\frac{1}{\sin\theta}\hat{{\bf \phi}}\nonumber
\end{eqnarray}
where $\hat{{\bf \theta}}$ and $\hat{{\bf \phi}}$ are the standard unit basis vectors on the sphere and the other quantities defined above satisfy the following relations from differential geometry
\begin{eqnarray}
g_{ab}&=&\mathbf{e}_{a}\cdot\mathbf{e}_{b}\nonumber\\
g^{ab}&=&\mathbf{e}^{a}\cdot\mathbf{e}^{b}\nonumber
\end{eqnarray}
where $\{a,b\}$ index the $\theta$- and $\phi$-components. For future reference the gradient of a scalar and the covariant derivative of a vector field on the sphere in this notation read
\begin{eqnarray}
\nabla\psi&=&(\partial_{a}\psi)\hat{\mathbf{e}}^{a}\nonumber\\
\nabla_{b}v_{a}&=&\partial_{b}v_{a}-\Gamma^{c}_{ab}v_{c}\nonumber
\end{eqnarray}
where $\mathbf{v}$ is a vector field with components $\mathbf{v}=v_{\theta}\hat{\mathbf{e}}^{\theta}+v_{\phi}\hat{\mathbf{e}}^{\phi}$.  For completeness the expressions in component form for the first and second derivatives of a scalar function $\Psi$ are 
\begin{eqnarray}
\nabla_{\theta}\Psi&=&\partial_{\theta}\Psi\ \ \hat{\theta}\nonumber\\
\nabla_{\phi}\Psi&=&\frac{1}{\sin\theta}\partial_{\phi}\Psi\ \  \hat{\phi}\nonumber\\
\nabla_{\theta}\nabla_{\theta}\Psi&=&\partial^{2}_{\theta}\Psi\ \  \hat{\theta}\otimes\hat{\theta}\nonumber\\
\nabla_{\phi}\nabla_{\phi}\Psi&=&\frac{1}{\sin^{2}\theta}\partial^{2}_{\phi}\Psi
        +\frac{\cos\theta}{\sin\theta}\partial_{\theta}\Psi\ \  \hat{\phi}\otimes\hat{\phi}\nonumber\\
\nabla_{\phi}\nabla_{\theta}\Psi&=&\frac{1}{\sin\theta}\partial_{\theta}\partial_{\phi}\Psi
        -\frac{\cos\theta}{\sin^{2}\theta}\partial_{\phi}\Psi\ \  \hat{\phi}\otimes\hat{\theta}\nonumber\ .
\end{eqnarray}
In these expressions all derivatives in the $\phi$-direction have been scaled by $1/\sin\theta$ to account for the non-unit length of $\hat{\mathbf{e}}^{\phi}$.  Finally, for the 1-axis along the $\hat{\theta}$-direction and the 2-axis along the $\hat{\phi}$-direction, the magnitudes of the deflection and shear components in terms of the derivatives given above are 
\begin{eqnarray}
\alpha_{\theta}&=&-\nabla_{\theta}\Psi \nonumber\\
\alpha_{\phi}&=&-\nabla_{\phi}\Psi \nonumber\\
\gamma_{1}&=&-\frac{1}{2}\left(\nabla_{\phi}\nabla_{\phi}\Psi - \nabla_{\theta}\nabla_{\theta}\Psi\right)\nonumber\\
\gamma_{2}&=&\nabla_{\phi}\nabla_{\theta}\Psi\nonumber
\end{eqnarray}
 
The Poisson equation for a point particle at the top of the sphere reads
\begin{displaymath}
\nabla^{2}\Psi=\delta(\cos\theta-1)\delta(\phi)-\frac{1}{4\pi}
\end{displaymath}
where $\nabla^{2}$ is the Laplacian operator which for my choice of coordinates is
\begin{displaymath}
\nabla^{2} = \frac{1}{\sin\theta}\frac{\partial}{\partial\theta}\left(\sin\theta\frac{\partial}{\partial\theta}\right)+\frac{1}{\sin^{2}\theta}
\frac{\partial^{2}}{\partial\phi^{2}}\ .
\end{displaymath}
I have subtracted the mean density of the point particle on the sphere from the overall density in the Poisson equation above, anticipating the result that $\ell=0$ modes (i.e. constant modes) of the density field can be set to zero when solving for the potential. The solution to the Poisson equation above is
\begin{displaymath}
\Psi(\theta,\phi) = \frac{1}{4\pi}\log\left[1-\cos\theta\right]\ .
\end{displaymath}
This equation has no $\phi$ dependence because I have placed the point particle at the top of the sphere.  

Expressions for the deflection angle and shear due to a point mass can be obtained from the expressions for the gradient and the covariant derivative of a vector of the sphere given above. Since I will store the shear in an orthonormal basis aligned with $(\hat{\theta},\hat{\phi})$ on the sphere, the $\phi-\phi$ component of $\nabla_{a}\nabla_{b}\Psi(\theta,\phi)$ must be scaled by $1/\sin^{2}\theta$ in order to account for the non-unit length of $\hat{\mathbf{e}}^{\phi}$.  The deflection angle and shear components for a point mass at the top of the sphere are
\begin{eqnarray}
\alpha_{\theta}&=&-\frac{1}{4\pi}\frac{\sin\theta}{1-\cos\theta}\nonumber\\
\alpha_{\phi}&=&0\nonumber\\
\gamma_{1}&=&-\frac{1}{8\pi}\frac{1+\cos\theta}{1-\cos\theta}\nonumber\\
\gamma_{2}&=&0\nonumber
\end{eqnarray}
The results for the shear can be obtained from \citet{deputter2010} as well. Once the deflection angle and the shear for the point particle are known when the point particle is at the top of the sphere, it is straight forward to generalize the expressions for any two points on the sphere.  Given the point particle's position and the position under consideration, one simply computes the ``parallactic'' angle between the great circle connecting the point particle and the position under consideration and the $+Q$ axis of the local basis in which the shear is stored at each location on the sphere (i.e. $\hat\theta$).  This angle $\omega$ can then be used to rotate both the shear and deflection angle into the local basis as needed.

\section{The Parallel Transport of Tensors on the Sphere Along Geodesics}\label{app:paratrans}
In this section I discuss the parallel transport of tensors on the sphere.  The discussion in this section follows closely the notation of \citet{hobson2006}.  In general, the parallel transport of a tensor $t^{ab}$ along a curve $x^{c}(u)$ parametrized by an affine parameter $u$ is given by the solution of the following equation 
\begin{equation}
0=\frac{Dt^{ab}}{Du}=\frac{dt^{ab}}{du} + \Gamma_{dc}^{a}t^{db}\frac{dx^{c}}{du} + \Gamma_{dc}^{b}t^{ad}\frac{dx^{c}}{du}
\end{equation}
where the $\Gamma_{ab}^{c}$ are the Christoffel symbols and the notation $D/Du$ denotes the intrinsic derivative along the curve $x^{c}$.  The ray propagation algorithm defined above needs the parallel transport of the inverse magnification matrix (a tensor on the sphere) along great circle arcs (geodesics of the sphere) which connect each ray with its location on the next lens plane in angle.  

A simplified prescription for this operation can be defined as follows.  Consider the geodesic along the ``equator'' of the sphere at $\theta=\pi/2$.  Along this path I can write $\mathbf{x}(u)=(\pi/2,u)$ where I now have used the arc-length as $u$.  This parameter is affine for this geodesic since the tangent vector to the geodesic, $\hat{\mathbf{e}}^{\phi}$, is the same at all points along the geodesic.  I have neglected an over all shift in ``phase'' around the sphere so that the parallel transport starts at $u=u_{i}=0$ and ends at some $u=u_{f}$.  With this definition of the geodesic $x^{c}$, one sees that all of the Christoffel symbols from Appendix~\ref{app:lenspmass} vanish identically.  Thus parallel transport for any tensor along this geodesic is trivial; its components remain constant. For this geodesic also note that the tensor components are defined in a basis in which one of the basis vectors which is always aligned with the tangent vector of the geodesic.  

Using this construction, it is straight forward to generalize the parallel transport of any tensor between any two points on the sphere connected by a geodesic.  One simply first rotates the tensor components in the original coordinate system into a coordinate system where the geodesic is along the ``equator'' of the sphere.  Then the tensor is parallel transported along the geodesic in this new coordinate system, where its components remain unchanged during this operation.  Then at the final location of the tensor the components are rotated back into the old coordinate system, being careful to note that the tensor basis at the final point in the old coordinate system differs from the tensor basis at the initial point in the old coordinate system.  This procedure is exactly that described by \citet{challinor2002}, but the previous discussion has established that it applies not only to traceless, symmetric polarization tensors, but to all tensors defined on the surface of the sphere which are parallel transported along geodesics.  

In practice I use the following equivalent prescription for parallel transporting tensors along geodesics on the sphere.  I simply perform a three-dimensional rotation of the local tensor basis 3-vectors about the center of the sphere.  The axis and angle of this rotation is fixed by requiring them to trace the geodesic path used for parallel transporting.  Then using the rotated tensor basis vectors and those at the final location of the sphere, I compute the angle by which to rotate the tensor components.  Finally using this angle, I express the tensor components in this final basis. Note that this procedure applies just as well to vectors on the surface of the sphere by the same argument as given above.  

\section{Equivalence of the Recurrence Relations for the Inverse Magnification Matrix}\label{app:rayprop}
The recurrence relation for the inverse magnification matrix obtained directly from the discretization of the lensing equations (see Equation~\ref{eqn:discsheartraj})
\begin{equation}\label{eqn:Atrajold}
\mathbf{A}_{k} = I - \sum_{i}^{k-1}\frac{D_{ik}}{D_{k}}\mathbf{U}_{i}\mathbf{A}_{i}
\end{equation} 
where $\mathbf{A}_{k}$ is the inverse magnification matrix for the rays at the $k$-th lensing plane, $\mathbf{U}_{i}$ is the matrix of second partial derivatives of the lensing potential at the $i$-th plane, $D_{k}\equiv r(\chi_{k})$, and $D_{ik} \equiv r(\chi_{k}-\chi_{i})$.  I have switched to matrix notation here with bold symbols denoting matrices.  With this definition of $D_{ik}$ one can verify the identity
\begin{equation}\label{eqn:angdistID}
D_{ac}=\frac{D_{a}D_{bc}+D_{ab}D_{c}}{D_{b}}
\end{equation}
directly from the properties of the sine and hyperbolic sine functions. Additionally in this notation $D_{ac}=-D_{ca}$.

With this identity in hand, I can now verify that Equation~(\ref{eqn:Atraj}) is equivalent to Equation~(\ref{eqn:Atrajold}) using induction (see \citeauthor{seitz1992} \citeyear{seitz1992} for a similar proof).  Once can verify directly that up to $k=2$ the two relations are the same.  Now assume they are equivalent up to $k-1$.  Then for $\mathbf{A}_{k}$ I get by substituting Equation~(\ref{eqn:Atrajold}) into Equation~(\ref{eqn:Atraj})
\begin{eqnarray}
\mathbf{A}_{k}&=&\left(1-\frac{D_{k-1}}{D_{k}}\frac{D_{k-2,k}}{D_{k-2,k-1}}\right)\mathbf{A}_{k-2}\nonumber\\
&& + \frac{D_{k-1}}{D_{k}}\frac{D_{k-2,k}}{D_{k-2,k-1}}\mathbf{A}_{k-1} - \frac{D_{k-1,k}}{D_{k}}\mathbf{U}_{k-1}\mathbf{A}_{k-1}\nonumber\\
&=&I-\frac{D_{k-1,k}}{D_{k}}\mathbf{U}_{k-1}\mathbf{A}_{k-1} - \frac{D_{k-2,k}}{D_{k}}\mathbf{U}_{k-2}\mathbf{A}_{k-2} \nonumber\\
&&+ \sum_{i=1}^{k-3}\left[\frac{D_{k-1}}{D_{k}}\frac{D_{k-2,k}}{D_{k-2,k-1}}\left(\frac{D_{i,k-2}}{D_{k-2}} -\frac{D_{i,k-1}}{{D_{k-1}}}\right)\right. \nonumber\\
&&\ \ \ \ \ \ \ \ \ \ \ \ \ \ \ \ - \left.\frac{D_{i,k-2}}{D_{k-2}}\right]\mathbf{U}_{i}\mathbf{A}_{i}\ .
\end{eqnarray}
By applying the identity in Equation~(\ref{eqn:angdistID}) twice to the expression in the brackets, I get
\begin{eqnarray}
\lefteqn{\frac{D_{k-1}}{D_{k}}\frac{D_{k-2,k}}{D_{k-2,k-1}}\left(\frac{D_{i,k-2}}{D_{k-2}}-\frac{D_{i,k-1}}{{D_{k-1}}}\right) - \frac{D_{i,k-2}}{D_{k-2}}}&&\nonumber\\
&&=\frac{D_{k-1}}{D_{k}}\frac{D_{k-2,k}}{D_{k-2,k-1}}\left(\frac{D_{k-1}D_{i,k-2}+D_{k-1,i}D_{k-2}}{D_{k-1}D_{k-2}}\right) \nonumber\\
&&\ \ \ \ \ \ - \frac{D_{i,k-2}}
{D_{k-2}}\nonumber\\
&&=-\frac{D_{k-2,k}D_{i}}{D_{k}D_{k-2}}-\frac{D_{i,k-2}}{D_{k-2}}\nonumber\\
&&=-\frac{D_{i}D_{k-2,k} + D_{i,k-2}D_{k}}{D_{k}D_{k-2}}\nonumber\\
&&=-\frac{D_{ik}}{D_{k}}\nonumber\ .
\end{eqnarray}
Thus by induction Equations~(\ref{eqn:Atraj}) and (\ref{eqn:Atrajold}) are equivalent.

\section{Radial Shear Interpolation for Galaxy Images}\label{app:galradint}
In this appendix I verify that the radial shear interpolation for finding galaxy images presented in Section~\ref{sec:galimages} is equivalent to properly computing the inverse magnification matrix for the galaxy from the previous two lens planes.  Assume a galaxy with comoving distance $\chi_{g}$ falls in lens plane $n+1$ so that $\chi_{n+1-1/2}<\chi_{g}<\chi_{n+1+1/2}$.  Then the correct equations to compute the inverse magnification matrix at the galaxy's radial location is (cf. Equation~\ref{eqn:Atraj}) 
\begin{eqnarray}\label{eqn:truegatraj}
\mathbf{A}_{g}&=&\left(1-\frac{D_{n}}{D_{g}}\frac{D_{n-1,g}}{D_{n-1,n}}\right)\mathbf{A}_{n-1}+\frac{D_{n}}{D_{g}}\frac{D_{n-1,g}}{D_{n-1,n}}\mathbf{A}_{n}\nonumber\\
&&-\frac{D_{n,g}}{D_{g}}\mathbf{U}_{n}\mathbf{A}_{n}\ ,
\end{eqnarray}
where I have now switched to the notation of Appendix~\ref{app:rayprop}. For practical reasons, CALCLENS actually implements the radial interpolation using the following equation
\begin{eqnarray}\label{eqn:gatraj}
\mathbf{A}_{g}&=&\left(1-\frac{D_{n+1}}{D_{g}}\frac{D_{n,g}}{D_{n,n+1}}\right)\mathbf{A}_{n}\nonumber\\
&&+\frac{D_{n+1}}{D_{g}}\frac{D_{n,g}}{D_{n,n+1}}\mathbf{A}_{n+1}\ .
\end{eqnarray}
This last equation matches directly the procedure described in Section~\ref{sec:galimages}. By direct substitution of Equation~(\ref{eqn:Atraj}) into Equation~(\ref{eqn:gatraj}), one can verify that the relation given in Equation~(\ref{eqn:truegatraj}) is in fact equivalent to that in Equation~(\ref{eqn:gatraj}). This derivation requires the identity given in Equation~(\ref{eqn:angdistID}) and also the following identity
\begin{equation}
D_{ad}=\frac{D_{ac}D_{bd}-D_{ab}D_{cd}}{D_{bc}}\ ,
\end{equation}
which again can be verified directly from the properties of the sine and hyperbolic sine functions.

\section{The Deflection Angle and Shear for the Epanechnikov Kernel on the Sphere}\label{sec:epkern}
I use the results of \citet{deputter2010} to compute the deflection angle and shear of the Epanechnikov kernel.  \citet{deputter2010} derived the full-sky relationship between the angle-averaged convergence and the angle-averaged tangential shear,
\begin{equation}\label{eqn:dss}
\langle\gamma_{T}\rangle = 2\frac{\cos\theta}{1+\cos\theta}\bar{\kappa}(<\theta) - \langle\kappa\rangle(\theta)
\end{equation}
where one is averaging around a point at the top of the sphere and 
\begin{equation}
\bar{\kappa}(<\theta)= \frac{1}{2\pi(1-\cos\theta)}\int_{0}^{2\pi}\int_{0}^{\theta}d\phi\,d\theta'\,\sin\theta'\kappa(\theta',\phi)
\end{equation}
\begin{equation}\label{eqn:aav}
\langle\kappa\rangle(\theta) = \frac{1}{2\pi}\int_{0}^{2\pi}d\phi\,\kappa(\theta',\phi)\ .
\end{equation}
In order to compute the deflection angle, I use the following relationship from \citet{deputter2010}
\begin{equation}\label{eqn:asp}
\bar{\kappa}(<\theta) = \frac{\sin\theta}{2(1-\cos\theta)}\partial_{\theta}\langle\Psi\rangle
\end{equation}
where $\partial_{\theta}\langle\Psi\rangle$ is the angle-averaged partial derivative of the lensing potential $\Psi$, similar to Equation~(\ref{eqn:aav}).

There are however a few important details to consider before completing the calculation.  First, \citet{deputter2010} use the standard definition of the convergence which differs from this work by a factor of two (i.e., $\kappa\rightarrow\kappa/2$  in Equations~(\ref{eqn:dss}) and (\ref{eqn:asp}) given above).  Second, one expects that at scales greater than smoothing length of the kernel, the solution for the deflection angle and shear should equal that of a point mass derived above. In order for the formula derived by \citet{deputter2010} to reproduce the expected large scale limit, one must subtract the mean density of the kernel on the sphere, $1/4\pi$.  Finally, the Epanechnikov kernel is symmetric in $\phi$ so that the results from \citet{deputter2010} can be used to compute quantities which are not angle averaged.  Thus the equation being solved is
\begin{equation}
\nabla^{2}\Psi = K(\theta;\,\sigma) - \frac{1}{4\pi}\ .
\end{equation}

With these details in mind, the deflection angle and shear components for an Epanechnikov kernel placed at the top of the sphere are
\begin{eqnarray}
\alpha_{\theta} &=&
\left\{\begin{array}{ll}
-h(\theta;\,\sigma)\frac{1-\cos\theta}{\sin\theta} & \theta < \sigma\\
-\frac{1}{4\pi}\frac{\sin\theta}{1-\cos\theta} & \theta \geq \sigma
\end{array}\right.\\
\alpha_{\phi} &=&  0\\
\gamma_{1} &=&  \left\{\begin{array}{ll}
-\frac{1}{2}\left(2\frac{\cos\theta}{1+\cos\theta}h(\theta;\,\sigma)\right. &\\
\ \ \ \ \ \ \ \ \ \ \ \ \ \ \ \ \ \ \ \ \ \ \ \ \ \ \ \ \ \  \left.- K(\theta;\,\sigma)+\frac{1}{4\pi}\right) & \theta < \sigma\\
-\frac{1}{8\pi}\frac{1+\cos\theta}{1-\cos\theta} & \theta \geq \sigma
\end{array}\right.\\
\gamma_{2} &=&  0\ 
\end{eqnarray}
where $h(\theta;\,\sigma)$ is
\begin{eqnarray}
\lefteqn{h(\theta;\,\sigma)=\frac{1}{{\cal N}(\sigma)(1-\cos\theta)}\times}&&\\
&& \left[\left(\frac{\theta}{\sigma}\right)^{2}\left(-2\sinc(\theta/\pi) + \cos\theta + \sinc(\theta/2/\pi)^{2}\right) \right.\nonumber\\
&& \ \ \ \ \ \ \ \ \ + 1 - \cos\theta \bigg] - \frac{1}{4\pi}\ .
\end{eqnarray}



\bsp
\label{lastpage}

\end{document}